\documentclass[11pt]{article}
\usepackage[linktocpage]{hyperref}
\usepackage{latexsym,graphicx,mathrsfs,amsfonts,subfig}
\usepackage{slashed, color}
\usepackage{multirow}
\usepackage{tikz}
\usepackage{empheq}
\usepackage{amsmath}
\allowdisplaybreaks[2]
\usepackage[multiple]{footmisc}%for multiple footnotes
\usepackage{textcomp}
\usepackage{amssymb}%for semidirect product symbol
\usetikzlibrary{positioning}
\usepackage{amscd}
\usepackage{amsthm}
\usepackage{array} 

\renewcommand{\thefootnote}{\fnsymbol{footnote}}
\numberwithin{equation}{section}
\usepackage{dsfont}
\usepackage[toc,page]{appendix}

\newcommand*\widefbox[1]{\fbox{\hspace{2em}#1\hspace{2em}}}%Used together with the empheq package

\DeclareFontFamily{U}{MnSymbolC}{}
\DeclareSymbolFont{MnSyC}{U}{MnSymbolC}{m}{n}
\DeclareFontShape{U}{MnSymbolC}{m}{n}{
	<-6>  MnSymbolC5
	<6-7>  MnSymbolC6
	<7-8>  MnSymbolC7
	<8-9>  MnSymbolC8
	<9-10> MnSymbolC9
	<10-12> MnSymbolC10
	<12->   MnSymbolC12}{}
\DeclareMathSymbol{\intprod}{\mathbin}{MnSyC}{'270}
\newcommand{\ov}{\overline}

\newcommand{\del}{\partial}

\newcommand{\til}{\widetilde}

\let\nc\newcommand
\let\renc\renewcommand
\nc{\wbar}{\overline}
\let\td\tilde
\let\wtd\widetilde
\let\wht\widehat
\let\mcl\mathcal

\nc{\ab}{{\bar{a}}} \nc{\at}{\tilde{a}} \nc{\ah}{\hat{a}}
\nc{\bb}{{\bar{b}}} 
\nc{\bh}{\hat{b}}
\nc{\cb}{{\bar{c}}} \nc{\ct}{\tilde{c}} %\nc{\ch}{\hat{c}}
\nc{\db}{{\bar{d}}} \nc{\dt}{\tilde{d}} \renc{\dh}{\hat{d}}
\nc{\eb}{{\bar{e}}} \nc{\et}{\tilde{e}} \nc{\eh}{\hat{e}}
\nc{\fb}{{\bar{f}}} \nc{\ft}{\tilde{f}} \nc{\fh}{\hat{f}}
\nc{\ib}{{\bar{\imath}}} \nc{\ih}{\hat{\imath}} %\nc{\it}{\tilde{\imath}}
\nc{\jb}{{\bar{\jmath}}} \nc{\jt}{\tilde{\jmath}} \nc{\jh}{\hat{\jmath}}
\nc{\kb}{{\bar{k}}} \nc{\kt}{\tilde{k}} \nc{\kh}{\hat{k}}
\nc{\lb}{{\bar{l}}} \nc{\lt}{\tilde{l}} \nc{\lh}{\hat{l}}
\nc{\mb}{{\bar{m}}} \nc{\mt}{\tilde{m}} \nc{\mh}{\hat{m}}
\nc{\nb}{{\bar{n}}} \nc{\nt}{\tilde{n}} \nc{\nh}{\hat{n}}
\nc{\ob}{{\bar{o}}} \nc{\ot}{\tilde{o}} \nc{\oh}{\hat{o}}
\nc{\pb}{{\bar{p}}} \nc{\pt}{\tilde{p}} \nc{\ph}{\hat{p}}
\nc{\qb}{{\bar{q}}} \nc{\qt}{\tilde{q}} \nc{\qh}{\hat{q}}
\nc{\rb}{{\bar{r}}} \nc{\rt}{\tilde{r}} \nc{\rh}{\hat{r}}
\renc{\sb}{{\bar{s}}} \nc{\st}{\tilde{s}} \nc{\sh}{\hat{s}}
\nc{\tb}{{\bar{t}}} \renc{\th}{\hat{t}} %\nc{\tt}{\tilde{t}}
\nc{\ub}{{\bar{u}}} \nc{\ut}{\tilde{u}} \nc{\uh}{\hat{u}}
\nc{\vb}{{\bar{v}}} \nc{\vt}{\tilde{v}} \nc{\vh}{\hat{v}}
\nc{\wt}{\tilde{w}} \nc{\wh}{\hat{w}}
\nc{\xb}{{\bar{x}}} \nc{\xt}{\tilde{x}} \nc{\xh}{\hat{x}}
\nc{\yb}{{\bar{y}}} \nc{\yt}{\tilde{y}} \nc{\yh}{\hat{y}}
\nc{\zb}{{\bar{z}}} \nc{\zt}{\tilde{z}} 
\nc{\Ab}{\wbar{A}} \nc{\At}{\wtd{A}} \nc{\Ah}{\wht{A}}
\nc{\Bb}{\wbar{B}} \nc{\Bt}{\wtd{B}} \nc{\Bh}{\wht{B}}
\nc{\Cb}{\wbar{C}} \nc{\Ct}{\wtd{C}} \nc{\Ch}{\wht{C}}
\nc{\Db}{\wbar{D}} \nc{\Dt}{\wtd{D}} \nc{\Dh}{\wht{D}}
\nc{\Eb}{\wbar{E}} \nc{\Et}{\wtd{E}} \nc{\Eh}{\wht{E}}
\nc{\Fb}{\wbar{F}} \nc{\Ft}{\wtd{F}} \nc{\Fh}{\wht{F}}
\nc{\Gb}{\wbar{G}} \nc{\Gt}{\wtd{G}} \nc{\Gh}{\wht{G}}
\nc{\Hb}{\wbar{H}} \nc{\Ht}{\wtd{H}} \nc{\Hh}{\wht{H}}
\nc{\Ib}{\wbar{I}} \nc{\It}{\wtd{I}} \nc{\Ih}{\wht{I}}
\nc{\Jb}{\wbar{J}} \nc{\Jt}{\wtd{J}} \nc{\Jh}{\wht{J}}
\nc{\Kb}{\wbar{K}} \nc{\Kt}{\wtd{K}} \nc{\Kh}{\wht{K}}
\nc{\Lb}{\wbar{L}} \nc{\Lt}{\wtd{L}} \nc{\Lh}{\wht{L}}
\nc{\Mb}{\wbar{M}} \nc{\Mt}{\wtd{M}} \nc{\Mh}{\wht{M}}
\nc{\Nb}{\wbar{N}} \nc{\Nt}{\wtd{N}} \nc{\Nh}{\wht{N}}
\nc{\Ob}{\wbar{O}} \nc{\Ot}{\wtd{O}} \nc{\Oh}{\wht{O}}
\nc{\Pb}{\wbar{P}} \nc{\Pt}{\wtd{P}} \nc{\Ph}{\wht{P}}
\nc{\Qb}{\wbar{Q}} \nc{\Qt}{\wtd{Q}} \nc{\Qh}{\wht{Q}}
\nc{\Rb}{\wbar{R}} \nc{\Rt}{\wtd{R}} \nc{\Rh}{\wht{R}}
\nc{\Sb}{\wbar{S}} \nc{\St}{\wtd{S}} \nc{\Sh}{\wht{S}}
\nc{\Tb}{\wbar{T}} \nc{\Tt}{\wtd{T}} \nc{\Th}{\wht{T}}
\nc{\Ub}{\wbar{U}} \nc{\Ut}{\wtd{U}} \nc{\Uh}{\wht{U}}
\nc{\Vb}{\wbar{V}} \nc{\Vt}{\wtd{V}} \nc{\Vh}{\wht{V}}
\nc{\Wb}{\wbar{W}} \nc{\Wt}{\wtd{W}} \nc{\Wh}{\wht{W}}
\nc{\Xb}{\wbar{X}} \nc{\Xt}{\wtd{X}} \nc{\Xh}{\wht{X}}
\nc{\Yb}{\wbar{Y}} \nc{\Yt}{\wtd{Y}} \nc{\Yh}{\wht{Y}}
\nc{\Zb}{\wbar{Z}} \nc{\Zt}{\wtd{Z}} \nc{\Zh}{\wht{Z}}

\nc{\CA}{\mcl{A}} \nc{\CAb}{\wbar{\CA}} \nc{\CAt}{\wtd{\CA}} \nc{\CAh}{\wht{\CA}}
\nc{\CB}{\mcl{B}} \nc{\CBb}{\wbar{\CB}} \nc{\CBt}{\wtd{\CB}} \nc{\CBh}{\wht{\CB}}
\nc{\CC}{\mcl{C}} \nc{\CCb}{\wbar{\CC}} \nc{\CCt}{\wtd{\CC}} \nc{\CCh}{\wht{\CC}}
\nc{\cDt}{\wtd{\cC}} \nc{\cDh}{\wht{\cD}}
\nc{\CE}{\mcl{E}} \nc{\CEb}{\wbar{\CE}} \nc{\CEt}{\wtd{\CE}} \nc{\CEh}{\wht{\CE}}
\nc{\CF}{\mcl{F}} \nc{\CFb}{\wbar{\CF}} \nc{\CFt}{\wtd{\CF}} \nc{\CFh}{\wht{\CF}}
\nc{\CG}{\mcl{G}} \nc{\CGb}{\wbar{\CG}} \nc{\CGt}{\wtd{\CG}} \nc{\CGh}{\wht{\CG}}
\nc{\CH}{\mcl{H}} \nc{\CHb}{\wbar{\CH}} \nc{\CHt}{\wtd{\CH}} \nc{\CHh}{\wht{\CH}}
\nc{\CI}{\mcl{I}} \nc{\CIb}{\wbar{\CI}} \nc{\CIt}{\wtd{\CI}} \nc{\CIh}{\wht{\CI}}
\nc{\CJ}{\mcl{J}} \nc{\CJb}{\wbar{\CJ}} \nc{\CJt}{\wtd{\CJ}} \nc{\CJh}{\wht{\CJ}}
\nc{\CK}{\mcl{K}} \nc{\CKb}{\wbar{\CK}} \nc{\CKt}{\wtd{\CK}} \nc{\CKh}{\wht{\CK}}
\nc{\CL}{\mcl{L}} \nc{\CLb}{\wbar{\CL}} \nc{\CLt}{\wtd{\CL}} \nc{\CLh}{\wht{\CL}}
\nc{\CM}{\mcl{M}} \nc{\CMb}{\wbar{\CM}} \nc{\CMt}{\wtd{\CM}} \nc{\CMh}{\wht{\CM}}
\nc{\CN}{\mcl{N}} \nc{\CNb}{\wbar{\CN}} \nc{\CNt}{\wtd{\CN}} \nc{\CNh}{\wht{\CN}}
\nc{\CO}{\mcl{O}} \nc{\COb}{\wbar{\CO}} \nc{\COt}{\wtd{\CO}} \nc{\COh}{\wht{\CO}}
\nc{\CQ}{\mcl{Q}} \nc{\CQb}{\wbar{\CQ}} \nc{\CQt}{\wtd{\CQ}} \nc{\CQh}{\wht{\CQ}}
\nc{\CR}{\mcl{R}} \nc{\CRb}{\wbar{\CR}} \nc{\CRt}{\wtd{\CR}} \nc{\CRh}{\wht{\CR}}
\nc{\CS}{\mcl{S}} \nc{\CSb}{\wbar{\CS}} \nc{\CSt}{\wtd{\CS}} \nc{\CSh}{\wht{\CS}}
\nc{\CT}{\mcl{T}} \nc{\CTb}{\wbar{\CT}} \nc{\CTt}{\wtd{\CT}} \nc{\CTh}{\wht{\CT}}
\nc{\CU}{\mcl{U}} \nc{\CUb}{\wbar{\CU}} \nc{\CUt}{\wtd{\CU}} \nc{\CUh}{\wht{\CU}}
\nc{\CV}{\mcl{V}} \nc{\CVb}{\wbar{\CV}} \nc{\CVt}{\wtd{\CV}} \nc{\CVh}{\wht{\CV}}
\nc{\CW}{\mcl{W}} \nc{\CWb}{\wbar{\CW}} \nc{\CWt}{\wtd{\CW}} \nc{\CWh}{\wht{\CW}}
\nc{\CX}{\mcl{X}} \nc{\CXb}{\wbar{\CX}} \nc{\CXt}{\wtd{\CX}} \nc{\CXh}{\wht{\CX}}
\nc{\CY}{\mcl{Y}} \nc{\CYb}{\wbar{\CY}} \nc{\CYt}{\wtd{\CY}} \nc{\CYh}{\wht{\CY}}
\nc{\CZ}{\mcl{Z}} \nc{\CZb}{\wbar{\CZ}} \nc{\CZt}{\wtd{\CZ}} \nc{\CZh}{\wht{\CZ}}

\let\eps\epsilon
\let\ups\upsilon
\let\veps\varepsilon
\let\vtht\vartheta
\let\vsgm\varsigma
\let\vphi\varphi
\let\vrho\varrho

\nc{\alphab}{\bar{\alpha}} \nc{\alphat}{\td{\alpha}} \nc{\alphah}{\hat{\alpha}}
\nc{\betab}{\bar{\beta}}   \nc{\betat}{\td{\beta}}   \nc{\betah}{\hat{\beta}} 
\nc{\gammab}{\bar{\gamma}} \nc{\gammat}{\td{\gamma}} \nc{\gammah}{\hat{\gamma}} 
\nc{\deltab}{\bar{\delta}} \nc{\deltat}{\td{\delta}} \nc{\deltah}{\hat{\delta}} 
\nc{\epsilonb}{\bar{\eps}} \nc{\epsilont}{\td{\eps}} \nc{\epsilonh}{\hat{\eps}} 
\nc{\vepsb}{\bar{\veps}}   \nc{\vepst}{\td{\veps}}   \nc{\vepsh}{\hat{\veps}} 
\nc{\zetab}{\bar{\zeta}}   \nc{\zetat}{\td{\zeta}}   \nc{\zetah}{\hat{\zeta}} 
\nc{\etab}{\bar{\eta}}     
\nc{\etah}{\hat{\eta}} 
\nc{\thetab}{\bar{\theta}} \nc{\thetat}{\td{\theta}} \nc{\thetah}{\hat{\theta}} 
\nc{\vthetab}{\bar{\vtht}} \nc{\vthetat}{\td{\vtht}} \nc{\vthetah}{\hat{\vtht}} 
\nc{\lambdat}{\td{\lambda}} \nc{\lambdah}{\hat{\lambda}} 
\nc{\iotab}{\bar{\iota}}   \nc{\iotat}{\td{\iota}}   \nc{\iotah}{\hat{\iota}} 
\nc{\kappab}{\bar{\kappa}} \nc{\kappat}{\td{\kappa}} \nc{\kappah}{\hat{\kappa}} 
\nc{\lmdb}{\bar{\lmd}}     \nc{\lmdt}{\td{\lmd}}     \nc{\lmdh}{\hat{\lmd}} 
\nc{\mub}{\bar{\mu}}       \nc{\mut}{\td{\mu}}       \nc{\muh}{\hat{\mu}} 
\nc{\nub}{\bar{\nu}}       \nc{\nut}{\td{\nu}}       \nc{\nuh}{\hat{\nu}} 
\nc{\xib}{\bar{\xi}}       \nc{\xit}{\td{\xi}}       \nc{\xih}{\hat{\xi}} 
\nc{\pib}{\bar{\pi}}       \nc{\pit}{\td{\pi}}       \nc{\pih}{\hat{\pi}} 
\nc{\vpib}{\bar{\vpi}}     \nc{\vpit}{\td{\vpi}}     \nc{\vpih}{\hat{\vpi}} 
\nc{\rhob}{\bar{\rho}}     \nc{\rhot}{\td{\rho}}     \nc{\rhoh}{\hat{\rho}} 
\nc{\vrhob}{\bar{\vrho}}   \nc{\vrhot}{\td{\vrho}}   \nc{\vrhoh}{\hat{\vrho}} 
\nc{\sigmab}{\bar{\sigma}} \nc{\sigmat}{\td{\sigma}} \nc{\sigmah}{\hat{\sigma}} 
\nc{\vsigmab}{\bar{\vsgm}} \nc{\vsigmat}{\td{\vsgm}} \nc{\vsigmah}{\hat{\vsgm}} 
\nc{\taub}{\bar{\tau}}     \nc{\taut}{\td{\tau}}     \nc{\tauh}{\hat{\tau}} 
\nc{\upsb}{\bar{\ups}} \nc{\upst}{\td{\ups}} \nc{\upsh}{\hat{\ups}} 
\nc{\phib}{\bar{\phi}}     \nc{\phit}{\td{\phi}}     \nc{\phih}{\hat{\phi}} 
\nc{\varphib}{\bar{\vphi}}   \nc{\varphit}{\td{\vphi}}   \nc{\varphih}{\hat{\vphi}} 
\nc{\chib}{\bar{\chi}}     
\nc{\chih}{\hat{\chi}} 
\nc{\psib}{\bar{\psi}}     
\nc{\psih}{\hat{\psi}} 
\nc{\omegab}{\bar{\omega}} \nc{\omegat}{\td{\omega}} \nc{\omegah}{\hat{\omega}} 

\nc{\Gammab}{\wbar{\Gamma}}     \nc{\Gammat}{\wtd{\Gamma}}     \nc{\Gammah}{\wht{\Gamma}}
\nc{\Deltab}{\wbar{\Delta}}     \nc{\Deltat}{\wtd{\Delta}}     \nc{\Deltah}{\wht{\Delta}}
\nc{\Thetab}{\wbar{\Theta}}     \nc{\Thetat}{\wtd{\Theta}}     \nc{\Thetah}{\wht{\Theta}}
\nc{\Lambdab}{\wbar{\Lambda}}   \nc{\Lambdat}{\wtd{\Lambda}}   \nc{\Lambdah}{\wht{\Lambda}}
\nc{\Xib}{\wbar{\Xi}}           \nc{\Xit}{\wtd{\Xi}}           \nc{\Xih}{\wht{\Xi}}
\nc{\Pib}{\wbar{\Pi}}           \nc{\Pit}{\wtd{\Pi}}           \nc{\Pih}{\wht{\Pi}}
\nc{\Sigmab}{\wbar{\Sigma}}     \nc{\Sigmat}{\wtd{\Sigma}}     \nc{\Sigmah}{\wht{\Sigma}}
\nc{\Upsilonb}{\wbar{\Upsilon}} \nc{\Upsilont}{\wtd{\Upsilon}} \nc{\Upsilonh}{\wht{\Upsilon}}
\nc{\Phib}{\wbar{\Phi}}         \nc{\Phit}{\wtd{\Phi}}         \nc{\Phih}{\wht{\Phi}}
\nc{\Psib}{\wbar{\Psi}}         \nc{\Psit}{\wtd{\Psi}}         \nc{\Psih}{\wht{\Psi}}
\nc{\Omegab}{\wbar{\Omega}}     \nc{\Omegat}{\wtd{\Omega}}     \nc{\Omegah}{\wht{\Omega}}
\nc{\Varepsilon}{\mathcal{E}}

\newcommand{\cD}{{\cal D}}

\nc{\balpha}{\bar{\alpha}}
\nc{\bbeta}{\bar{\beta}}
\nc{\bgamma}{\bar{\gamma}}
\nc{\bm}{\bar{m}}
\nc{\bn}{\bar{n}}
\nc{\bp}{\bar{p}}
\nc{\al}{\alpha}
\nc{\bt}{\beta}
\nc{\gm}{\gamma}
\nc{\zh}{\wht{z}}
\nc{\zhb}{\ov{\wht{z}}}
\nc{\mbh}{\wht{\ov{m}}}
\nc{\bc}{|_{x^2=0}}

\nc{\tal}{\til{\al}}
\nc{\tbt}{\til{\bt}}
\nc{\tgm}{\til{\gm}}

\nc{\wb}{\ov{w}}
\nc{\teta}{\til{\eta}}
\nc{\tpsi}{\til{\psi}}

\def\IL{\relax{\rm I\kern-.18em L}}
\def\IH{\relax{\rm I\kern-.18em H}}
\def\IB{\relax{\rm I\kern-.18em B}}
\def\ID{\relax{\rm I\kern-.18em D}}
\def\IE{\relax{\rm I\kern-.18em E}}
\def\IF{\relax{\rm I\kern-.18em F}}

\def\IG{\relax\hbox{$\inbar\kern-.3em{\rm G}$}}
\def\IGa{\relax\hbox{${\rm I}\kern-.18em\Gamma$}}
\def\IH{\relax{\rm I\kern-.18em H}}
\def\II{\relax{\rm I\kern-.18em I}}
\def\IK{\relax{\rm I\kern-.18em K}}
\def\IP{\relax{\rm I\kern-.18em P}}
\def\IQ{\relax\hbox{$\inbar\kern-.3em{\rm Q}$}}

\def\hat{\widehat}
\def\CM {{\cal M}}
\def\CN {{\cal N}}
\def\CR {{\cal R}}

\def\CF {{\cal F}}
\def\CJ {{\cal J}}

\def\CL {{\cal L}}
\def\CV {{\cal V}}
\def\CO {{\cal O}}
\def\CZ {{\cal Z}}
\def\CE {{\cal E}}
\def\CG {{\cal G}}
\def\CH {{\cal H}}
\def\CC {{\cal C}}
\def\CB {{\cal B}}
\def\CS {{\cal S}}
\def\CA{{\cal A}}
\def\CK{{\cal K}}
\def\CQ{{\cal Q}}

\def\p{\partial}
\def\pb{{\bar \p}}

\def\vt#1#2#3{ {\vartheta[{#1 \atop  #2}](#3\vert \tau)} }

\def\jb{{\bar j}}

\def\inbar{\,\vrule height1.5ex width.4pt depth0pt}

\nc{\hTheta}{\hat{\Theta}}
\nc{\vp}{\varphi}
\nc{\tg}{\widetilde{g}}

\let\OLDthebibliography\thebibliography
\renewcommand\thebibliography[1]{
	\OLDthebibliography{#1}
	\setlength{\parskip}{5pt}
	\setlength{\itemsep}{0pt plus 0.3ex}
}
\usepackage{titlesec}
\titleformat*{\section}{\bfseries\large}
\usepackage{tikz}                                       %
\usepackage{tikz-cd}                                    %
\usetikzlibrary{cd}                                     %
\usetikzlibrary{calc}                                   %
\usepackage{lipsum}
\usepackage[sorting=none]{biblatex}                     % 
\usepackage{hyperref}                                   %
\bibliography{references}                               %
\begin{document}
\addtolength{\baselineskip}{1.5mm}

\thispagestyle{empty}

\vbox{}
\vspace{3.0cm}

\begin{center}
	\centerline{\LARGE{Topological-Holomorphic ${\cal N}=4$ Gauge Theory: From Langlands Duality of}}
	\bigskip
	\centerline{\LARGE{Holomorphic Invariants to Mirror Symmetry of Quasi-topological Strings}} 

\vspace{2.5cm}

	{Zhi-Cong~Ong\footnote{E-mail: zc\textunderscore ong@nus.edu.sg} and Meng-Chwan~Tan\footnote{E-mail: mctan@nus.edu.sg}}
	\\[2mm]
	{\it Department of Physics\\
		National University of Singapore \\%\\[1mm]
		2 Science Drive 3, Singapore 117551} \\[1mm] 
\end{center}

\vspace{2.0cm}

\centerline{\bf Abstract}\smallskip \noindent

We perform a topological-holomorphic twist of $\mathcal{N}=4$ supersymmetric gauge theory on a four-manifold of the form $M_4=\Sigma_1 \times \Sigma_2$ with Riemann surfaces $\Sigma_{1,2}$, and unravel the mathematical implications of its physics. In particular,  we consider different linear combinations of the resulting scalar supercharges under $S$-duality, where this will allow us to derive novel topological and holomorphic invariants of $M_4$ and their Langlands duals. As the twisted theory can be topological along $\Sigma_1$ whence we can dimensionally reduce it to 2d, via the effective sigma-model on $\Sigma_2$, we can also relate these 4d invariants and their Langlands duals to the mirror symmetry of Higgs bundles and that of quasi-topological strings described by the sheaf of chiral differential operators. As an offshoot, we would be able to obtain a fundamental understanding from 4d gauge theory, why chiral differential operators are purely perturbative objects.    

\newpage

\renewcommand{\thefootnote}{\arabic{footnote}}
\setcounter{footnote}{0}

\tableofcontents
\section{Introduction, Summary and Acknowledgements}

\bigskip\noindent\textit{Introduction}
\vspace*{0.5em}\\
We perform a topological-holomorphic twist of $\mathcal{N}=4$ SYM theory on a four-manifold $M_4=\Sigma_{1} \times \Sigma_{2}$ with real, simple and compact gauge group $G$, where $\Sigma_{1,2}$ are Riemann surfaces. 
%Twisting generally gives us more than one scalar supercharge in the theory. 
In this topological-holomorphic twist, we end up with four scalar supercharges, where the cohomology of linear combinations of these scalar supercharges gives either a theory that is topological on all of $M_4=\Sigma_{1} \times \Sigma_{2}$, or topological on $\Sigma_{1}$ and holomorphic on $\Sigma_{2}$, or  holomorphic on both $\Sigma_{1}$ and $\Sigma_{2}$. 

We aim to study the mathematical implications of these different combinations via $S$-duality of ${\cal N} = 4$ gauge theory, and by establishing dualities between the theories in four and two dimensions through the technique of dimensional reduction used in \cite{bershadsky1995topological}, where shrinking the topological directions of $M_4$ along $\Sigma_1$ in the first two cases gives us an effective sigma-model that is either topological or holomorphic on $\Sigma_{2}$. In particular, the holomorphic sigma-model is a quasi-topological sigma-model with local operators described by the sheaf of chiral differential operators (CDO)~\cite{gorbounov1999gerbes, gorbounov2000gerbes, gorbounov2003gerbes} on target space, and it defines a quasi-topological string. 

Note that a topological-holomorphic twist in 4d gauge theory was first considered in \cite{kapustin2006holomorphic}, where the theory studied was $\mathcal{N}=2$ SYM with matter. Here, besides studying an $\mathcal{N}=4$ SYM theory, we will also examine different linear combinations of scalar supercharges that have no analogs in \cite{kapustin2006holomorphic}.

Let us now give a brief plan and summary of the paper.

\bigskip\noindent\textit{A Brief Plan and Summary of the Paper}
\vspace*{0.5em}\\
In $\S$\ref{section: top-hol twist}, we perform a topological-holomorphic twist of $\mathcal{N}=4$ SYM on $M_4 = \Sigma_1 \times \Sigma_1$, where $\Sigma_{1,2}$ are closed Riemann surfaces. Such a twist is achieved by embedding three of the $U(1)_R$ subgroups of the $SU(4)_R$ $R$-symmetry of the theory in the $U(1)_E$ holonomy groups of both $\Sigma_{1,2}$. Four scalar supercharges are obtained after the twist, and cohomologies of different linear combinations of these supercharges give the spectrum of either topological, topological-holomorphic, or holomorphic theories.

In $\S$\ref{section: 4d topological theory}, we consider the cohomology of  a linear combination $\mathcal Q$ of supercharges that gives a topological theory on $M_4$. \emph{Novel} topological invariants of this theory are correlation functions of operators $\mathcal{O}$ in the $\mathcal{Q}$-cohomology, and they are of the form
\begin{equation}\label{4d topo corr function 0}
    \langle \Pi_i\mathcal{O}^{(m)}_i \rangle_{4d}\big(\tau, G\big)= \int_{\mathcal{M}}D\phi\; \Pi_i\mathcal{O}^{(m)}_i e^{-S}
\end{equation}
in \eqref{4d topo corr function}, where $D\phi$ represents the integration measure over all field configurations, and $\mathcal{M}$ is the \emph{novel} moduli space of field configurations that this theory localizes on according to its BPS equations in \ref{4d BPS}.

In $\S$\ref{section: topological A-model 2d}, we perform a dimensional reduction by compactifying the topological theory of  $\S$\ref{section: 4d topological theory} along $\Sigma_1$, where we consider $\Sigma_1$ to be closed of genus $g\geq 2$. We arrive at an $A$-model in complex structure $I$ on $\Sigma_2$ with $\mathcal{N}=(4,4)$ supersymmetry and target space $\mathcal{M}_H^G(\Sigma_1)$, the moduli space of Hitchin's equations on $\Sigma_1$. In complex structure $I$, $\mathcal{M}_H^G(\Sigma_1)$ can be identified with $\mathcal{M}_{\text{Higgs}}^G(\Sigma_1)$, the moduli space of stable Higgs $G$-bundles on $\Sigma_1$. Correlation functions are Gromov-Witten invariants of $\mathcal{M}_H^G(\Sigma_1)$, and topological invariance of the theory allows us to have a 4d-2d correspondence 
\begin{equation}\label{4d-2d corr top 0}
   \boxed{ \langle \Pi_i\mathcal{O}^{(m)}_i \rangle_{4d}\big(\tau, G\big) = \langle \Pi_i\tilde{\mathcal{O}}^{(p)}_i \rangle_{2d} \big(\tau, \mathcal{M}^G_{\text{Higgs}}(\Sigma_{1})\big)}
\end{equation}
in \eqref{4d-2d corr top}.

In $\S$\ref{section: 4d top-hol theory}, we consider a different cohomology by taking a different linear combination $\mathcal{Q}'$ of scalar supercharges from before, giving us a theory that is topological along $\Sigma_1$ and holomorphic along $\Sigma_2$, which coordinates are $(z, \bar{z})$ and $(w, \bar{w})$, respectively. Correlation functions of local operators $\mathcal{O}'$ in the $\mathcal{Q}'$-cohomology now have a holomorphic dependence on the coordinates of $\Sigma_2$, where we have a \emph{novel} holomorphic invariant on $M_4$ 
\begin{equation}\label{4d top-hol corr function 0}
    \langle \Pi_i\mathcal{O}'_i \rangle_{4d}\big(w, G\big) = \int_{\mathcal{M}'}D\phi'\; \Pi_i\mathcal{O}'_i e^{-S}
\end{equation}
in \eqref{4d top-hol corr function}, associated with the \emph{novel} moduli space $\mathcal M'$ defined by \eqref{4d BPS top-hol}. 

In $\S$\ref{section: holomoprhic sigma model 2d}, we perform a dimensional reduction by compactifying the topological-holomorphic theory of $\S$\ref{section: 4d top-hol theory} along $\Sigma_1$, where we consider $\Sigma_1$ to be closed of genus $g\geq 2$. We now obtain a holomorphic or quasi-topological sigma-model in complex structure $J+\alpha K$ (where $\alpha \in \mathbb{C}$) on $\Sigma_2$ with $\mathcal{N}=(4,4)$ supersymmetry and target space $\mathcal{M}_{\text{flat}}^{G_{\mathbb{C}}}(\Sigma_1)$, the moduli space of flat complexified connections on $\Sigma_1$. Observables are local holomorphic operators dependent only on $w$, which belong to the \u{C}ech cohomology $H^*_{\text{\u{C}ech}}(\mathcal{M}_{\text{flat}}^{G_{\mathbb{C}}}(\Sigma_1), \Omega_{\text{cdo}})$, where $\Omega_{\text{cdo}}$ is the sheaf  of CDOs on $\mathcal{M}_{\text{flat}}^{G_{\mathbb{C}}}(\Sigma_1)$. Correlation functions are evaluations over $\mathcal{M}_{\text{flat}}^{G_{\mathbb{C}}}(\Sigma_1)$ of a product of classes in $H^*_{\text{\u{C}ech}}(\mathcal{M}_{\text{flat}}^{G_{\mathbb{C}}}(\Sigma_1), \Omega_{\text{cdo}})$ that correspond to these local holomorphic operators.
Topological invariance of the theory along $\Sigma_1$ implies a 4d-2d correspondence of correlation functions and thus invariants, such that we will have 
\begin{equation}\label{4d-2d top-hol corr 0}
    \boxed{ \langle \Pi_i\mathcal{O}'_i \rangle_{4d}\big(w, G\big)= \int_{\mathcal{M}_{\text{flat}}^{G_{\mathbb{C}}}(\Sigma_1)} \bigotimes_i H^*_{\text{\u{C}ech}, i}  =  \text{CDO}\big(w, \mathcal{M}^{G_{\mathbb{C}}}_{\text{flat}}(\Sigma_{1})\big)}
\end{equation}
in \eqref{4d-2d top-hol corr}. As there are no non-perturbative instanton contributions to the path integral on the RHS  of \eqref{4d top-hol corr function 0}, the RHS of \eqref{4d-2d top-hol corr 0} which descends from it, would not have any non-perturbative worldsheet instanton contributions either. Thus, we have a fundamental understanding from 4d gauge theory, why CDOs are purely perturbative objects. 

In $\S$\ref{section: holomorphic m4}, we consider only \emph{one} of the scalar supercharges in $\mathcal{Q}'$-cohomology, where we obtain a theory that is fully-holomorphic on $M_4$ i.e., the theory is holomorphic on both $\Sigma_1$ and $\Sigma_2$. Correlation functions now have a dependence on both $z$ and $w$, where we have a \emph{novel} fully-holomorphic invariant on $M_4$ given by 
\begin{equation}\label{4d top-hol corr function single scalar charge 0}
    \langle \Pi_i\mathcal{O}'_{i} \rangle_{4d,0}\big(w,z, \tau, G\big) = \int_{\mathcal{M}_{0}'}D\phi'\; \Pi_i\mathcal{O}'_{0,i} e^{-S}
\end{equation}
in \eqref{4d top-hol corr function single scalar charge}, associated with the \emph{novel} moduli space $\mathcal M'_0$ defined by \eqref{4d bps single scalar charge 1}.

In $\S$\ref{section: langlands duality}, we show that $S$-duality of the above-mentioned theories result in a Langlands duality of the topological and holomorphic invariants, and mirror symmetry. Specifically, in $\mathcal{Q}$-cohomology, we have, in \eqref{langlands dual 4d top corr},
\begin{equation}\label{langlands dual 4d top corr 0}
   \boxed{ \langle \Pi_i\mathcal{O}^{(m)}_i \rangle_{4d}\big(\tau, G\big) \longleftrightarrow \langle \Pi_i\mathcal{O}^{(m)}_i \rangle_{4d}\big(-1/n_{\mathfrak{g}}\tau, {^LG}\big)}
\end{equation}
a Langlands duality of the 4d topological invariants. In \eqref{langlands dual 2d higgs}, we have
\begin{equation}\label{langlands dual 2d higgs 0}
   \boxed{  \langle \Pi_i\tilde{\mathcal{O}}^{(p)}_i \rangle_{2d} \big(\tau, \mathcal{M}^G_{\text{Higgs}}(\Sigma_{1})\big) \longleftrightarrow \langle \Pi_i\tilde{\mathcal{O}}^{(p)}_i \rangle_{2d} \big(-1/n_{\mathfrak{g}}\tau, \mathcal{M}^{^LG}_{\text{Higgs}}(\Sigma_{1})\big)}
\end{equation}
which can be interpreted as a mirror symmetry of Higgs bundles. 

If $\Sigma_2=\mathbb{R} \times I$ where $I$ is an interval, we have an open $A$-model and a  homological mirror symmetry of the $\tau$-dependent category of $A$-branes
\begin{equation}\label{dual Cat A brane 0}
    \boxed{\text{Cat}_{\text{$A$-branes}}\big(\tau, \mathcal{M}^{G}_{\text{Higgs}}(\Sigma_1)\big) \longleftrightarrow
    \text{Cat}_{\text{$A$-branes}}\big(-1/n_{\mathfrak{g}}\tau, \,\mathcal{M}^{^LG}_{\text{Higgs}}(\Sigma_1)\big) }
\end{equation}
in \eqref{dual Cat A brane}. 
%where $\mathcal{M}^{G}_{\text{Higgs}}$ and $\mathcal{M}^{^LG}_{\text{Higgs}}$ are mirror manifolds. 

In $\mathcal{Q}'$-cohomology, we have, in \eqref{langlands dual 4d top-hol corr},
\begin{equation}\label{langlands dual 4d top-hol corr 0}
   \boxed{ \langle \Pi_i\mathcal{O}'_i \rangle_{4d}\big(w, G\big) \longleftrightarrow \langle \Pi_i\mathcal{O}'_i \rangle_{4d}\big(w, {^LG}\big)}
\end{equation}
a Langlands duality of the 4d holomorphic invariants. 
In \eqref{langlands dual 2d top-hol }, we have 
\begin{equation}\label{langlands dual 2d top-hol 0}
\boxed{ \text{CDO}\big(w, \mathcal{M}^{G_{\mathbb{C}}}_{\text{flat}}(\Sigma_{1})\big)\longleftrightarrow\text{CDO}\big(w, \mathcal{M}^{{^LG}_{\mathbb{C}}}_{\text{flat}}(\Sigma_{1})\big)}
\end{equation}
a mirror symmetry between CDOs and thus quasi-topological strings on $\mathcal{M}^{G_{\mathbb{C}}}_{\text{flat}}(\Sigma_{1})$ and its mirror $\mathcal{M}^{{^LG}_{\mathbb{C}}}_{\text{flat}}(\Sigma_{1})$. 
If we consider only one of the scalar supercharges in $\mathcal{Q}'$-cohomology, we have, in \eqref{langlands 4d hol single},
\begin{equation}\label{langlands 4d hol single 0}
\boxed{\langle \Pi_i\mathcal{O}'_{i} \rangle_{4d,0}\big(w,z, \tau, G\big) \longleftrightarrow \langle \Pi_i\mathcal{O}'_{i} \rangle_{4d,0}\big(w,z, -1/n_{\mathfrak{g}}\tau, {^LG}\big)}
\end{equation}
a Langlands duality of the 4d fully-holomorphic invariants.

In $\S$\ref{section: web}, we will present a novel web of mathematical relations, summarizing the dualities, correspondences, and identifications between the various novel mathematical objects we physically derived in $\S$\ref{section: 4d topological theory}--\ref{section: holomorphic m4}. 

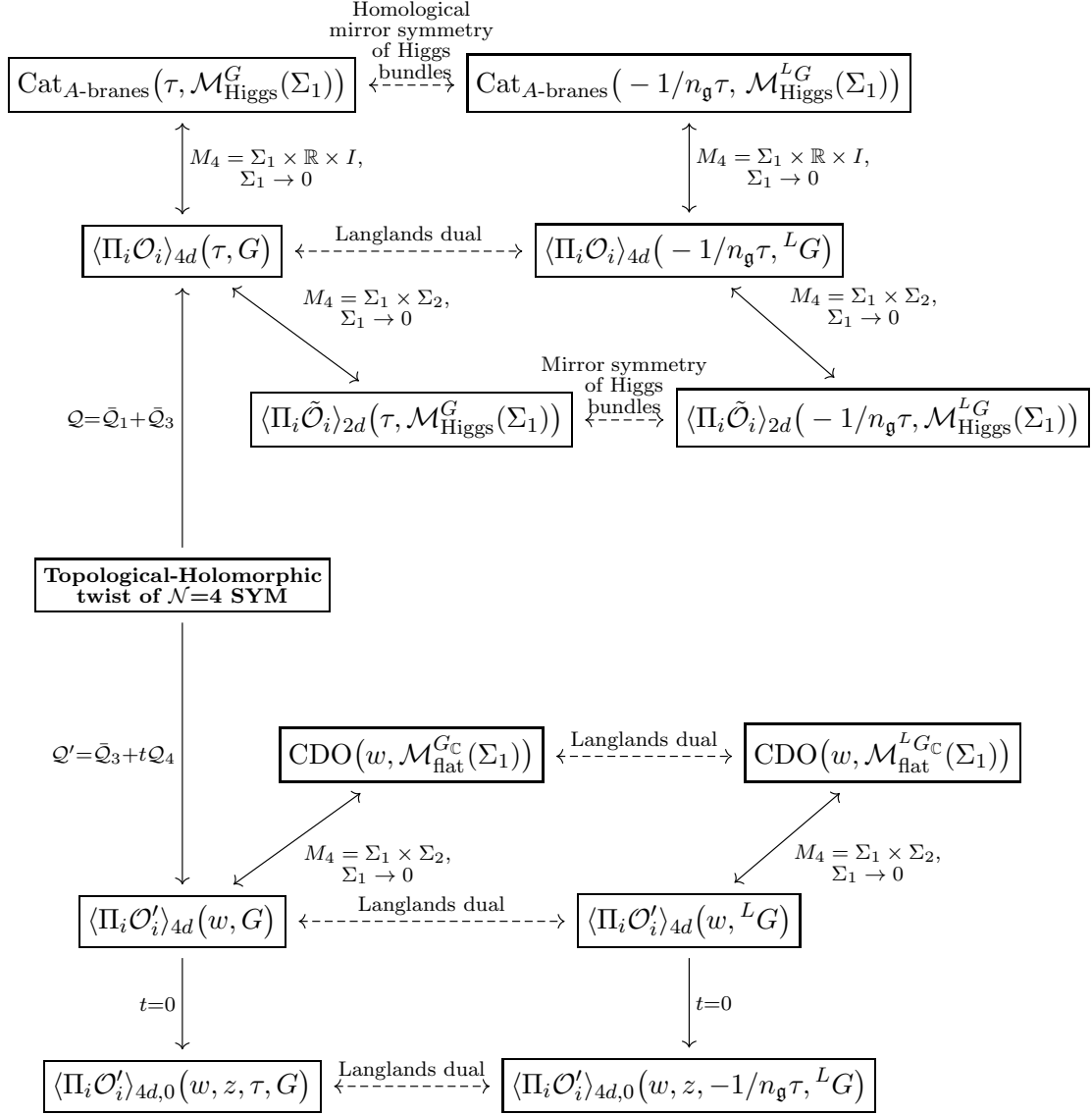
\begin{figure}
\begin{center}
  \begin{tikzcd}[column sep=-50pt,row sep=35pt]
  \boxed{\text{Cat}_{\text{$A$-branes}}\big(\tau, \mathcal{M}^{G}_{\text{Higgs}}(\Sigma_1)\big)}
  \arrow[rr, leftrightarrow, dashed, "\substack{\text{Homological}\\\text{ mirror symmetry }\\\text{of Higgs}\\\text{ bundles}}"]
  \arrow[d, leftrightarrow,  "\substack{\text{$M_4=\Sigma_{1} \times \mathbb{R}\times I$,}\\\text{$\Sigma_1\to 0$}}"]
  &&\boxed{\text{Cat}_{\text{$A$-branes}}\big(-1/n_{\mathfrak{g}}\tau, \,\mathcal{M}^{^LG}_{\text{Higgs}}(\Sigma_1)\big)}
  \arrow[d, leftrightarrow,  "\substack{\text{$M_4=\Sigma_{1} \times \mathbb{R}\times I$,}\\\text{$\Sigma_1\to 0$}}"]\\
  \boxed{\langle \Pi_i\mathcal{O}_i \rangle_{4d}\big(\tau, G\big)}
  \arrow[dr, leftrightarrow, "\substack{\text{$M_4=\Sigma_{1} \times \Sigma_2$,}\\\text{$\Sigma_1\to 0$}}"]
  \arrow[rr, leftrightarrow, dashed, "\text{Langlands dual}"]
  &&\boxed{\langle \Pi_i\mathcal{O}_i \rangle_{4d}\big(-1/n_{\mathfrak{g}}\tau, {^LG}\big)}
  \arrow[drr, leftrightarrow,  "\substack{\text{$M_4=\Sigma_{1} \times \Sigma_2$,}\\\text{$\Sigma_1\to 0$}}"]\\
  &\boxed{ \langle \Pi_i\tilde{\mathcal{O}}_i \rangle_{2d} \big(\tau, \mathcal{M}^G_{\text{Higgs}}(\Sigma_{1})\big)}
  \arrow[rrr, leftrightarrow, dashed, "\substack{\text{Mirror symmetry}\\\text{of Higgs}\\\text{bundles}}"]
  &&&\boxed{\langle \Pi_i\tilde{\mathcal{O}}_i \rangle_{2d} \big(-1/n_{\mathfrak{g}}\tau, \mathcal{M}^{^LG}_{\text{Higgs}}(\Sigma_{1})\big)}\\
  %%%%%%%%%%%%%%%%%%%%%%%%%%%%%%%%%%%%%%%%%%%%%%%%%%%%%%%%%%%%%%%%%%%%%%%%%%%%%%%%%%%%%
  \boxed{\substack{\textbf{Topological-Holomorphic}\\\textbf{twist of $\mathcal{N}$=4 SYM} }}
  \arrow[uu, "\mathcal{Q}=\bar{\mathcal{Q}}_1+\bar{\mathcal{Q}}_3" ]
  \arrow[dd, "\mathcal{Q}'=\bar{\mathcal{Q}}_3+t\mathcal{Q}_4"']\\
  %%%%%%%%%%%%%%%%%%%%%%%%%%%%%%%%%%%%%%%%%%%%%%%%%%%%%%%%%%%%%%%%%%%%%%%%%%%%%%%%%%%%%%%
   &\boxed{ \text{CDO}\big(w, \mathcal{M}^{G_{\mathbb{C}}}_{\text{flat}}(\Sigma_{1})\big)}
  \arrow[rrr, leftrightarrow, dashed, "\text{Langlands dual}"]
  &&&\boxed{\text{CDO}\big(w, \mathcal{M}^{{^LG}_{\mathbb{C}}}_{\text{flat}}(\Sigma_{1})\big)}\\ 
  \boxed{\langle \Pi_i\mathcal{O}'_i \rangle_{4d}\big(w, G\big)}
  \arrow[rr, leftrightarrow, dashed, "\text{Langlands dual}"]
  \arrow[ur, leftrightarrow, "\substack{\text{$M_4=\Sigma_{1} \times \Sigma_2$,}\\\text{$\Sigma_1\to 0$}}"']
  \arrow[d, "t=0"']
  &&\boxed{\langle \Pi_i\mathcal{O}'_i \rangle_{4d}\big(w, {^LG}\big)}
  \arrow[urr, leftrightarrow, "\substack{\text{$M_4=\Sigma_{1} \times \Sigma_2$,}\\\text{$\Sigma_1\to 0$}}"']
  \arrow[d,   "t=0"]\\
  \boxed{\langle \Pi_i\mathcal{O}'_i \rangle_{4d,0}\big(w,z,\tau, G\big)}
  \arrow[rr, leftrightarrow, dashed, "\text{Langlands dual}"]
  &&\boxed{\langle \Pi_i\mathcal{O}'_i \rangle_{4d,0}\big(w,z,-1/n_{\mathfrak{g}}\tau, {^LG}\big)}\\
  %\boxed{ \text{CDO}\big(w, \tau, \mathcal{M}^G_{\text{Higgs}}(\Sigma_{1})\big)}
  %\arrow[u, leftrightarrow , "\substack{\text{$M_4=\Sigma_{1} \times \Sigma_2$,}\\\text{$\Sigma_1\to 0$}}"]
  %\arrow[rr, leftrightarrow, dashed, "\substack{\text{Mirror symmetry}\\\text{of Higgs}\\\text{bundles}}"]
  %&&\boxed{\text{CDO} \big(w, -1/n_{\mathfrak{g}}\tau, \mathcal{M}^{^LG}_{\text{Higgs}}(\Sigma_{1})\big)}
  %\arrow[u, leftrightarrow , "\substack{\text{$M_4=\Sigma_{1} \times \Sigma_2$,}\\\text{$\Sigma_1\to 0$}}"']\\
    \end{tikzcd}
\caption{A novel web of mathematical relations stemming from a topological-holomorphic twist of an $\mathcal{N}=4$ supersymmetric gauge theory on $M_4$} \label{fig: web 0}
\end{center}
\end{figure}

In summary, the physics of the topological-holomorphic twist of $\mathcal{N}=4$ supersymmetric gauge theory on $M_4$ gives us the novel web of mathematical relations shown in Fig.~\ref{fig: web 0}.

\bigskip\noindent\textit{Acknowledgements}
\vspace*{0.5em}\\
We would like to thank M.~Ashwinkumar for many useful discussions. This work is supported in part by the MOE AcRF Tier 1 grant R-144-000-470-114. 

\section{The Topological-Holomorphic Twist}\label{section: top-hol twist}

{\renewcommand{\arraystretch}{1.4}
\begin{table}[ht]
\begin{center}
$\begin{tabular}{|c|c|c|c|c|c|c|c|c|} 
\hline
\text{Field} & $U(1)_{\Sigma_{1}}$ & $U(1)_{\Sigma_{2}}$ & $U(1)_X$ & $U(1)_Y$ & $U(1)_Z$ & $U(1)_{\Sigma_{1}}'$ & $U(1)_{\Sigma_{2}}'$ & Field (After) \\
\hline
$\lambda_{1\uparrow}$ & 1 & -1 & 1 & 1 & 0 & 2 & 0 & $\lambda_{1z}$\\
$\lambda_{2\uparrow}$ & 1 & -1 & 1 & -1 & 0 & 2 & -2 & $\lambda_{2\bar{w}z}$\\
$\lambda_{3\uparrow}$ & 1 & -1 & -1 & 0 & -1 & 0 & -2 & $\lambda_{3\bar{w}}$\\
$\lambda_{4\uparrow}$ & 1 & -1 & -1 & 0 & 1 & 0 & 0 & $\lambda_{4}$\\
\hline
$\lambda_{1\downarrow}$ & -1 & 1 & 1 & 1 & 0 & 0 & 2 & $\lambda_{1w}$\\
$\lambda_{2\downarrow}$ & -1 & 1 & 1 & -1 & 0 & 0 & 0 & $\lambda_{2}$\\
$\lambda_{3\downarrow}$ & -1 & 1 & -1 & 0 & -1 & -2 & 0 & $\lambda_{3\bar{z}}$\\
$\lambda_{4\downarrow}$ & -1 & 1 & -1 & 0 & 1 & -2 & 2 & $\lambda_{42\bar{z}}$\\
\hline
$\bar{\lambda}_{1\uparrow}$ & -1 & -1 & -1 & -1 & 0 & -2 & -2 & $\bar{\lambda}_{1\bar{w}\bar{z}}$\\
$\bar{\lambda}_{2\uparrow}$ & -1 & -1 & -1 & 1 & 0 & -2 & 0 & $\bar{\lambda}_{2\bar{z}}$\\
$\bar{\lambda}_{3\uparrow}$ & -1 & -1 & 1 & 0 & 1 & 0 & 0 & $\bar{\lambda}_{3}$\\
$\bar{\lambda}_{4\uparrow}$ & -1 & -1 & 1 & 0 & -1 & 0 & -2 & $\bar{\lambda}_{4\bar{w}}$\\
\hline
$\bar{\lambda}_{1\downarrow}$ & 1 & 1 & -1 & -1 & 0 & 0 & 0 & $\bar{\lambda}_{1}$\\
$\bar{\lambda}_{2\downarrow}$ & 1 & 1 & -1 & 1 & 0 & 0 & 2 & $\bar{\lambda}_{2w}$\\
$\bar{\lambda}_{3\downarrow}$ & 1 & 1 & 1 & 0 & 1 & 2 & 2 & $\bar{\lambda}_{3wz}$\\
$\bar{\lambda}_{4\downarrow}$ & 1 & 1 & 1 & 0 & -1 & 2 & 0 & $\bar{\lambda}_{4z}$\\
\hline
$B_1$ & 0 & 0 & 0 & -1 & -1 & 0 & -2 & $B_{\bar{w}}$\\
$B_2$ & 0 & 0 & 0 & -1 & 1 & 0 & 0 & $C$\\
$B_3$ & 0 & 0 & 2 & 0 & 0 & 2 & 0 & $B_{z}$\\
\hline
$B_1^{\dagger}$ & 0 & 0 & 0 & 1 & 1 & 0 & 2 & $B_{w}$\\
$B_2^{\dagger}$ & 0 & 0 & 0 & 1 & -1 & 0 & 0 & $C^{\dagger}$\\
$B_3^{\dagger}$ & 0 & 0 & -2 & 0 & 0 & -2 & 0 & $B_{\bar{z}}$\\
\hline
\end{tabular}$
\caption{Fields and their $U(1)_R$ charges. Up and down arrows in the subscript represent upper and lower components of the eight Weyl fermions in the 4d $\mathcal{N}=4$ theory. $U(1)_{\Sigma_{1}}'$ denotes the twisting of $\Sigma_{1}$ by $U(1)_X$ and $U(1)_{\Sigma_{2}}'$ denotes the twisting of $\Sigma_{2}$ by $U(1)_Y$ and $U(1)_Z$.}\label{chargetable}
\end{center}
\end{table}
}

In this section, we begin with a topological-holomorphic twist on a four-manifold $M_4=\Sigma_{1} \times \Sigma_{2}$, where three different $U(1)_R$ charges will be used to modify the spins of the fields, one along $\Sigma_{1}$ and two along $\Sigma_{2}$. The holonomy group of a Riemann surface is $U(1)_E$, and twisting on $\Sigma_{1}$ by a $U(1)_R$ $R$-symmetry amounts to modifying the spins of the fields along $\Sigma_{1,2}$.
We split the spinor indices of the fermions into those on $\Sigma_{1}$, indicated by a $\pm$, and those on $\Sigma_{2}$, indicated by $\tilde{\pm}$, respectively. For example, if a spinor field is (before twisting) written as $\Psi_{+\tilde{-}}$, it would mean that it is a section of $K_{\Sigma_{1}}^{1/2}\otimes K_{\Sigma_{2}}^{-1/2}$. Here, $K_{\Sigma_{1,2}}$ denotes the canonical line bundle of $\Sigma_{1,2}$.

Denote local complex coordinates on $\Sigma_{1}$ by $z$, $\bar{z}$, and those on $\Sigma_{2}$ by $w$, $\bar{w}$ with $z=x^1+ix^2$ and $w=x^3+ix^4$ ($x^1, x^2$ are Euclidean coordinates on $\Sigma_{1}$, while $x^3, x^4$ are the corresponding ones on $\Sigma_{2}$). 
We then perform a $U(1)$ twist of $\Sigma_{1}$ and $\Sigma_{2}$ by three different $U(1)_R$ $R$-symmetries, 
%subgroups of the $SU(4)_{\mathcal {R}}$ $R$-symmetry, 
labelled as $U(1)_X$, $U(1)_Y$ and $U(1)_Z$, whereby the corresponding charges of the fields are shown in Table \ref{chargetable}. 
$U(1)_{\Sigma_{1}}'$ denotes the twisting of $\Sigma_{1}$ by $U(1)_X$, and $U(1)_{\Sigma_{2}}'$ denotes the twisting of $\Sigma_{2}$ by $U(1)_Y$ and $U(1)_Z$.
%\footnote{The third $U(1)$ subgroup of $SU(4)_{\mathcal {R}}$ shall be labelled as $U(1)_Z$ but will not be further discussed in this paper, since it will not be used.}
Effectively, twisting shifts the spins of the fields by their $U(1)_R$ charges.  

For example, $(\lambda_{1\uparrow})_{+\tilde{-}}$ has $U(1)_X=1$ and $U(1)_Y=1$, so twisting effects
$(\lambda_{1\uparrow})_{+\tilde{-}}\to (\lambda_{1\uparrow})_{++,\tilde{-}\tilde{+}}=\lambda_{1z}.$ Here, we have used the fact that a field with indices $_{+-}$( $_{\tilde{+}\tilde{-}}$) transforms as a scalar on $\Sigma_{1}$($\Sigma_{2}$), $_{++}$($_{\tilde{+}\tilde{+}}$) as a $(1,0)$ form on $\Sigma_{1}$$(\Sigma_{2}$), and $_{--}$($_{\tilde{-}\tilde{-}}$) as a $(0,1)$ form on $\Sigma_{1}$($\Sigma_{2}$). 
After performing twistings of $(\lambda_{1\uparrow})_{+\tilde{-}}$ by $U(1)_X$, $U(1)_Y$ and $U(1)_Z$ on $\Sigma_{1}$ and $\Sigma_{2}$ as shown, $\lambda_{1\uparrow}$ becomes $\lambda_{1z}$, where it is now a section of $K_{\Sigma_{1}}\otimes \mathcal{O}_{\Sigma_{2}}$ (with $\mathcal{O}_{\Sigma_{2}}$ indicating that it is a scalar on $\Sigma_{2}$). This is reflected in Table \ref{chargetable}, together with the modified spins of the other fields in the theory. The last column reflects the geometrical properties of the fields on $M_4$ after twisting. The spins of all four components of the gauge field $A_{\mu}$ are unaffected by the twisting, as they have zero $U(1)_R$ charge. 
%studied in \cite{kapustin2006holomorphic}, where $\Sigma_{1}$ is a closed Riemann surface of genus $g$, and vice-versa for $C$. In particular, the twist we will be examining is the $\beta''$-twist as stated there. The model studied in \cite{kapustin2006holomorphic} is that of an $\mathcal{N}=2$ SYM theory coupled with a massless hypermultiplet in some representation $\mathcal{R}$. Taking $\mathcal{R}$ to be the adjoint representation brings us to the model studied in this section.
%Given that $M_4=\Sigma_{1} \times C$, where $\Sigma_{1}$ and $C$ are  Riemann surfaces, we have a block diagonal metric 
%\begin{equation}\label{4d metric}
%    g_{M_4} = \text{diag}\big(g_{\Sigma_{1}},  g_{C} \big),
%\end{equation}
%Twisting amounts to embedding the holonomoy group into the $R$-symmetry group. In our case, the holonomy group is $U(1)_{\Sigma_{1}}\times U(1)_{C}$ and an embedding is equivalent to identifying either of $U(1)_{\Sigma_{1}}$ or $U(1)_{C}$ with $U(1)_X$ or $U(1)_Y$. The choice made in this paper shall be to identify $U(1)_{\Sigma_{1}}$ with $U(1)_X$, and $U(1)_{C}$ with $U(1)_Y$.

With the modifications of the spins of the fields according to Table \ref{chargetable}, the effective action of the twisted $\mathcal{N}=4$ supersymmetric gauge theory on $M_4$ with  metric $ds^2=g_{z\bar{z}}dz\otimes d\bar{z}+g_{w\bar{w}}dw\otimes d\bar{w}$ is 
\begin{equation}\label{4d susy action}
    \begin{aligned}
    S &= \frac{1}{e^2}\int d^2z d^2w \sqrt{g}\,\text{Tr}\,\bigg[\frac{1}{2}F_{\mu\nu}F^{\mu\nu} + g^{z\bar{z}}D_{\mu}B_{\bar{z}}D^{\mu}B_{z}+D_{\mu}\tilde{C}D^{\mu}\tilde{C}^{\dagger}+D_{\mu}B_{\bar{w}}D^{\mu}B_{w}\\
    &-ig^{w\bar{w}}g^{z\bar{z}}\bar{\lambda}_{1\bar{w}\bar{z}}\big(D_{w}\lambda_{1z}+D_{z}\lambda_{1w}\big)
    -i\bar{\lambda}_{1}\big(D_{\bar{z}}\lambda_{1z}-g^{w\bar{w}}D_{\bar{w}}\lambda_{1w}\big)\\
    &-ig^{z\bar{z}}\bar{\lambda}_{2\bar{z}}\big(g^{w\bar{w}}D_{w}\lambda_{2\bar{w}z}+D_{z}\lambda_{2}\big)-ig^{w\bar{w}}\bar{\lambda}_{2w}\big(g^{z\bar{z}}D_{\bar{z}}\lambda_{2\bar{w}z}-D_{\bar{w}}\lambda_{2} \big)\\
    &-i\bar{\lambda}_{3}\big(g^{w\bar{w}}D_{w}\lambda_{3\bar{w}}+g^{z\bar{z}}D_{z}\lambda_{3\bar{z}} \big)-ig^{z\bar{z}}g^{w\bar{w}}\bar{\lambda}_{3wz}\big(D_{\bar{z}}\lambda_{3\bar{w}}-D_{\bar{w}}\lambda_{3\bar{z}} \big) \\
    &-ig^{w\bar{w}}\lambda_{4}\big(g^{w\bar{w}}D_{w}\bar{\lambda}_{4\bar{w}}+g^{z\bar{z}}D_{\bar{z}}\bar{\lambda}_{4z}\big) -ig^{z\bar{z}} g^{w\bar{w}}\lambda_{4w\bar{z}}\big(D_{z}\bar{\lambda}_{4\bar{w}} -D_{\bar{w}}\bar{\lambda}_{4z} \big)\\
    &+(g^{w\bar{w}})^2[B_{\bar{w}}, B_{w}]^2+[\tilde{C}, \tilde{C}^{\dagger}]^2 +(g^{z\bar{z}})^2[B_{z},B_{\bar{z}}]^2+ g^{w\bar{w}}[B_{\bar{w}}, B_{w}][\tilde{C}, \tilde{C}^{\dagger}]\\
    &+g^{z\bar{z}}[\tilde{C},\tilde{C}^{\dagger}][B_{z}, B_{\bar{z}}]+g^{w\bar{w}}g^{z\bar{z}}[B_{\bar{w}}, B_{w}][B_{z}, B_{\bar{z}}]\\
    &+ig^{w\bar{w}}B_{\bar{w}}\big(g^{z\bar{z}}\{\lambda_{4w\bar{z}}, \lambda_{1z}\} - \{\lambda_4 , \lambda_{1w} \}\big)- ig^{w\bar{w}}B_{w}\big(g^{z\bar{z}}\{\lambda_{2\bar{w}z}, \lambda_{3\bar{z}} \} - \{\lambda_{2}, \lambda_{3\bar{w}}  \} \big)\\
    &-iB_2\big(g^{w\bar{w}}g^{z\bar{z}}\{\lambda_{4w\bar{z}}, \lambda_{2\bar{w}z}\} - \{\lambda_4, \lambda_2 \}\big)-i\tilde{C}^{\dagger}\big(g^{w\bar{w}}\{\lambda_{1w},\lambda_{3\bar{w}} \} - g^{z\bar{z}}\{\lambda_{1z}, \lambda_{3\bar{z}}\} \big)\\
    &+ig^{z\bar{z}}B_{z}\big(g^{w\bar{w}}\{\lambda_{4w\bar{z}}, \lambda_{3\bar{w}}\} - \{\lambda_4, \lambda_{3\bar{z}}  \}\big)+ig^{z\bar{z}}B_{\bar{z}}\big(g^{w\bar{w}}\{\lambda_{1w}, \lambda_{2\bar{w}z}, \} - \{\lambda_{1z}, \lambda_2\} \big)\\
    &-ig^{w\bar{w}}B_{\bar{w}}\big(g^{z\bar{z}}\{\bar{\lambda}_{3wz}, \bar{\lambda}_{2\bar{z}} \} -\{\bar{\lambda}_{3}, \bar{\lambda}_{2w} \} \big) -ig^{w\bar{w}}B_{w}\big(g^{z\bar{z}}\{\bar{\lambda}_{1\bar{w}\bar{z}}, \bar{\lambda}_{4z} \} - \{\bar{\lambda}_{1},\bar{\lambda}_{4\bar{w}} \} \big)\\ &+i\tilde{C}\big(g^{w\bar{w}}g^{z\bar{z}}\{\bar{\lambda}_{3wz}, \bar{\lambda}_{1\bar{w}\bar{z}}  \} - \{\bar{\lambda}_{3},\bar{\lambda}_{1} \} \big) -i\tilde{C}^{\dagger}\big(g^{z\bar{z}}\{\bar{\lambda}_{2\bar{z}},\bar{\lambda}_{4z} \}+g^{w\bar{w}}\{\bar{\lambda}_{2w},\bar{\lambda}_{4\bar{w}} \} \big)\\
    &+ig^{z\bar{z}}B_{z}\big(g^{w\bar{w}}\{\bar{\lambda}_{2w}, \bar{\lambda}_{1\bar{w}\bar{z}}\} - \{\bar{\lambda}_{2\bar{z}}, \bar{\lambda}_{1} \} \big)-ig^{z\bar{z}}B_{\bar{z}}\big(g^{w\bar{w}}\{\bar{\lambda}_{3wz},\bar{\lambda}_{4\bar{w}} \} - \{\bar{\lambda}_{3},\bar{\lambda}_{4z} \}\big)\bigg]\\
    &-\frac{i\tau}{4\pi}\int_{M_4} \text{Tr}F \wedge F.
\end{aligned}
\end{equation}
where the last term is a topological term with a complex coupling parameter 
\begin{equation}
    \tau = \frac{\theta}{2\pi}+i\frac{4\pi}{e^2}.
\end{equation}

The supercharges are in the same representation as the fermions, and their spins are modified in the same manner as the fermion fields as shown in Table \ref{chargetable}. In particular, we first note that there are now four supercharges which are scalar in all four directions of $M_4$. Following the labelling of the fermion fields as in Table \ref{chargetable}, we will label them as  $\bar{\mathcal{Q}}_1$, $\mathcal{Q}_2$, $\bar{\mathcal{Q}}_3$, $\mathcal{Q}_4$. We first examine the supersymmetry algebra involving anti-commutators of the supercharges. They will be of the form 
\begin{equation}\label{susy identity}
    \{\mathcal{Q}_i, \bar{\mathcal{Q}}_j \}_{\mu} \propto \delta_{ij}P_{\mu},
\end{equation}
where $P_{\mu}$ is the four-momentum. We have the following anti-commutators of the scalar supercharges:
\begin{equation}
\begin{aligned}\label{susy algebra general}
    \{\bar{\mathcal{Q}}_1, \mathcal{Q}_{1z} \} &\propto P_z, &\quad \{\bar{\mathcal{Q}}_{2\bar{z}}, \mathcal{Q}_{2} \} &\propto P_{\bar{z}},\\
    \{\bar{\mathcal{Q}}_1, \mathcal{Q}_{1w} \} &\propto P_w, &\qquad \{\bar{\mathcal{Q}}_{2w}, \mathcal{Q}_{2} \} &\propto P_w,\\
    \{\bar{\mathcal{Q}}_3, \mathcal{Q}_{3\bar{w}} \} &\propto P_{\bar{w}}, &\qquad \{\bar{\mathcal{Q}}_{4z}, \mathcal{Q}_{4} \} &\propto P_z,\\
    \{\bar{\mathcal{Q}}_3, \mathcal{Q}_{3\bar{z}}\} &\propto P_{\bar{z}} &\qquad \{\bar{\mathcal{Q}}_{4\bar{w}}, \mathcal{Q}_{4} \} &\propto P_{\bar{w}}.
\end{aligned}
\end{equation}
 
 As mentioned, considering the cohomology of the different linear combinations of these supercharges will yield theories with different properties on $M_4$. The mathematical implications of these different combinations will be the subject of the following sections.

\section{A Topological Theory on $M_4$}\label{section: 4d topological theory}

We will first examine the $\mathcal{Q}$-cohomology, where
\begin{equation}\label{first top susy charge}
\mathcal{Q}=a\bar{\mathcal{Q}}_1+b\bar{\mathcal{Q}}_3,    
\end{equation}
and $a,b\in \mathbb{C}$. Here, we consider the case where $s=b/a\neq 0, \infty$.
Using the (twisted) supersymmetry relation \eqref{susy identity} and \eqref{susy algebra general}, we compute that
\begin{equation}
\begin{aligned}\label{susy algebra topological}
    \{\mathcal{Q}, \mathcal{Q}_{1z} \} &\propto P_z,\\
    \{\mathcal{Q}, \mathcal{Q}_{1w} \} &\propto P_w,\\
    \{\mathcal{Q}, \mathcal{Q}_{3\bar{w}} \} &\propto P_{\bar{w}},\\
    \{\mathcal{Q}, \mathcal{Q}_{3\bar{z}}\} &\propto P_{\bar{z}}.
\end{aligned}
\end{equation}
Since $P_{\mu}\propto \partial_{\mu}$, any possible dependence of correlation functions of the theory on spacetime coordinates will be trivial in $\mathcal{Q}$-cohomology. That is, for $s\neq 0, \infty$, the theory is topological on $M_4$ with respect to the $\mathcal{Q}$-cohomology.
%Nonetheless, since 
%we can still obtain a holomorphic theory by setting $s=0$ or $s=\infty$. However, because the theory is no longer topological, it is now unclear if we are able to perform dimensional reduction to obtain an effective sigma-model on one of the $\Sigma_{1}$. 

%The way around this is to write the action as an anti-commutator $S=\{\bar{\mathcal{Q}}_1, V \}$ (for $s=0$), where $V$ is a gauge fermion. This way, the theory will still be independent of the metric of both $\Sigma_{1}$, thus still allowing us to shrink either of them to obtain a sigma-model on the other. Crucially, the effective sigma-model is a holomorphic theory.

The transformations of the fields under $\mathcal{Q}$ %and $\bar{\mathcal{Q}}_3$ 
are
\begin{equation} \label{4d susy transformations}
    \begin{aligned}
        \delta A_{z}  &= -i\bar{\epsilon}_1 \lambda_{1z}\;, & \quad\quad \delta A_{\bar{z}}&= -i\bar{\epsilon}_3 \lambda_{3\bar{z}}\;,\\
        \delta A_w &= i\bar{\epsilon}_1 \lambda_{1w} \;,& \quad\quad\delta A_{\bar{w}}&= -i\bar{\epsilon}_3\lambda_{3\bar{w}}\;,\\
        \delta B_{\bar{w}} &=-\bar{\epsilon}_1\bar{\lambda}_{4\bar{w}}\;, & \quad\quad \delta B_{w}&=-\bar{\epsilon}_3\bar{\lambda}_{2w}\;,\\
        \delta C &=0\;, & \quad\quad  \delta C^{\dagger}&=-\bar{\epsilon}_3\bar{\lambda}_{1}+\bar{\epsilon}_1\bar{\lambda}_{3}\;,\\
        \delta B_{z} &=\bar{\epsilon}_3 \bar{\lambda}_{4z}\;, & \quad\quad\delta B_{\bar{z}}&=\bar{\epsilon}_1\bar{\lambda}_{2\bar{z}}\;,\\
        \delta \lambda_{1z}  &= -\bar{\epsilon}_3 D_{z}C \;, & \quad\quad
        \delta \lambda_{1w}  &=  -\bar{\epsilon}_3 D_{w}C\;,\\
        \delta \lambda_{2}  &= -i\bar{\epsilon}_1 D_{\bar{z}}B_{z}g^{z\bar{z}} + i\bar{\epsilon}_3 D_{w}B_{\bar{w}}g^{w\bar{w}}%[B_{z}, B_{\bar{z}}]+\bar{\epsilon} F_{z\bar{z}}-2i\bar{\bar{\epsilon}}
        \; & \quad\quad \delta \lambda_{2\bar{w}z}&=-i\bar{\epsilon}_1 D_{\bar{w}}B_{z} - i\bar{\epsilon}_3 D_{z}B_{\bar{w}} \;,\\%F_{\bar{w}z}-4i\bar{\bar{\epsilon}}D_{\bar{w}}B_{z}\;,\\
        %&-2\bar{\epsilon}(g^{w\bar{w}}F_{w\bar{w}}+i[B_{1\tilde{-}}, B^{\dagger}_{1\tilde{+}}]-i[B_{2\tilde{-}}, B^{\dagger}_{2\tilde{+}}])\;,\\
        \delta \lambda_{3\bar{w}}  &= -\bar{\epsilon}_1 D_{\bar{w}}C\;, & \quad\quad\delta \lambda_{3\bar{z}}&= -\bar{\epsilon}_1 D_{\bar{z}}C\;,\\
        %-4(\bar{\bar{\epsilon}}D_{\bar{z}}B^{\dagger}_{2\tilde{+}}-\bar{\epsilon} [B_{\bar{z}},B^{\dagger}_{2\tilde{+}}])\;,\\
        \delta \lambda_{4}  &= -i\bar{\epsilon}_{1} D_{\bar{w}}B_{w}g^{w\bar{w}}-i\bar{\epsilon}_{3} D_{z}B_{\bar{z}}g^{z\bar{z}}\;, & \quad\quad\delta \lambda_{4w\bar{z}}&=-i\bar{\epsilon}_{1} D_{\bar{z}}B_{w}+i\bar{\epsilon}_{3}D_{w}B_{\bar{z}} \;,\\
        \delta\bar{\lambda}_{1}  &= -\bar{\epsilon}_{1}(F_{z\bar{z}} -i[B_{z}, B_{\bar{z}}])g^{z\bar{z}}+i\bar{\epsilon}_{1}[C, C^{\dagger}]\; & \quad\quad\delta \bar{\lambda}_{1\bar{w}\bar{z}}&=-\bar{\epsilon}_{1} F_{\bar{w}\bar{z}}+i\bar{\epsilon}_{3} [B_{\bar{z}}, B_{\bar{w}}] \;,\\
        &-\bar{\epsilon}_{1}(F_{w\bar{w}}+i[B_{\bar{w}}, B_{w}])g^{w\bar{w}}\;,\\
        \delta\bar{\lambda}_{2\bar{z}}  &=\bar{\epsilon}_{3}[C, B_{\bar{z}}]\;, & \quad\quad
        \delta\bar{\lambda}_{2w} &= \bar{\epsilon}_{1}[C, B_{w}]\;,\\
        \delta\bar{\lambda}_{3} &=-\bar{\epsilon}_{3}(F_{z\bar{z}} -i[B_{z}, B_{\bar{z}}])g^{z\bar{z}}+i\bar{\epsilon}_{3}[C, C^{\dagger}]& \quad\quad\delta \bar{\lambda}_{3wz}&= -\bar{\epsilon}_{3} F_{wz}-i\bar{\epsilon}_{1} [B_{z}, B_{w}]\;,\\
        &-\bar{\epsilon}_{3}(F_{w\bar{w}}+i[B_{\bar{w}}, B_{w}])g^{w\bar{w}}\;,\\
        \delta \bar{\lambda}_{4\bar{w}} &=\bar{\epsilon}_{3}[C, B_{\bar{w}}]\;, & \quad\quad\delta \bar{\lambda}_{4z}&=\bar{\epsilon}_{1}[C, B_{z}]\;.
    \end{aligned}
\end{equation}
Modulo equations of motions, the variation gives 
\begin{equation}\label{nilpotent}
     \delta^2 \phi \propto [C, \phi],
\end{equation}
where $\phi$ represents any field in the theory, $C$ is a scalar in all directions on $M_4$, and \eqref{nilpotent} implies that $\mathcal{Q}$ is nilpotent up to gauge transformations. For convenience, we first set $s=1$ by letting $a= b =1$. 

We can introduce auxiliary fields $H$ and $P$ to modify the field transformations so that \eqref{nilpotent} holds without the use of equations of motion. The modified transformations under $\mathcal{Q}$ are
\begin{equation}\label{auxiliary transforms}
    \begin{aligned}
    \delta \bar{\lambda}_{1\bar{w}\bar{z}} &= H_{1\bar{w}\bar{z}}, &\qquad \delta \bar{\lambda}_1 &= P_1,\\
    \delta \lambda_{2\bar{w}z} &= H_{2\bar{w}z}, &\qquad \delta \lambda_2 &= P_2,\\
    \delta \bar{\lambda}_{3wz} &= H_{3wz}, &\qquad \delta \bar{\lambda}_3 &= P_3,\\
    \delta \lambda_{4w\bar{z}} &= H_{4w\bar{z}}, &\qquad \delta \lambda_4 &= P_4\\
    \end{aligned}
\end{equation}
where $\delta^2 H \propto [C, H]$ and $\delta^2 P \propto [C, P]$.

Letting
\begin{equation} 
    \begin{aligned}
    V_1 &= -\bar{\lambda}_{3wz}\bigg(\frac{1}{2}H_{1\bar{w}\bar{z}}-(F_{\bar{w}\bar{z}}-i [B_{\bar{z}}, B_{\bar{w}}])\bigg)g^{z\bar{z}}g^{w\bar{w}},\\
    V_2 &= \lambda_{4w\bar{z}}\bigg(\frac{1}{2}H_{2\bar{w}z}-i(D_{\bar{w}}B_{z} +  D_{z}B_{\bar{w}}) \bigg)g^{z\bar{z}}g^{w\bar{w}},\\
    V_3 &= -\bar{\lambda}_{1\bar{w}\bar{z}}\bigg(\frac{1}{2}H_{3wz}-(F_{wz}-i[B_{z}, B_{w}])\bigg)g^{z\bar{z}}g^{w\bar{w}},\\
    V_4 &= \lambda_{2\bar{w}z}\bigg(\frac{1}{2}H_{4w\bar{z}}-i(D_{\bar{z}}B_{w}-D_{z}B_{\bar{w}}) \bigg)g^{z\bar{z}}g^{w\bar{w}},\\
    V_5 &= -\lambda_{4}\bigg(\frac{1}{2}P_{2}-i(g^{z\bar{z}}D_{\bar{z}}B_{z}  - g^{w\bar{w}}D_{w}B_{\bar{w}}) \bigg),\\
    V_6 &= -\lambda_{2}\bigg(\frac{1}{2}P_{4}-i( g^{w\bar{w}}D_{\bar{w}}B_{w}+ g^{z\bar{z}}D_{z}B_{\bar{z}}) \bigg),\\
    V_7 &= (\bar{\lambda}_{3}+\bar{\lambda}_{1})\bigg(\frac{1}{2}P_{1}-(F_{z\bar{z}} -i[B_{z}, B_{\bar{z}}])g^{z\bar{z}}+i[C, C^{\dagger}]-(F_{w\bar{w}}+i[B_{\bar{w}}, B_{w}])g^{w\bar{w}} \bigg),\\
    V_8&= ig^{z\bar{z}}\lambda_{3\bar{z}}D_{z}C-ig^{w\bar{w}}\lambda_{1w}D_{\bar{w}}C,\\
    V_{9}&= ig^{w\bar{w}}\bar{\lambda}_{4\bar{w}}[C^{\dagger}, B_{w}]+ig^{z\bar{z}}\bar{\lambda}_{2\bar{z}}[C^{\dagger}, B_{z}],
    \end{aligned}
\end{equation}
the action can (up to suitable rescalings of the fields) be written in the following $\mathcal{Q}$-invariant form:
\begin{equation}\label{Q exact action}
    S = \frac{1}{e^2}\sum_i^{9}\int_{M_4}d^2z d^2w \sqrt{g}\, \, \text{Tr} \{ {\cal Q}, V_i \}  -\frac{i\tau}{4\pi}\int_{M_4} \text{Tr}F \wedge F .
\end{equation}
It is immediately observed that the action depends on the metric only through $\mathcal{Q}$-exact terms.  Consequently, the energy-momentum tensor will be of the form 
\begin{equation}\label{enery-momentum Q exact}
    T_{\mu\nu}=\frac{\delta S}{\delta g^{\mu\nu}}=\{ {\cal Q}, G_{\mu\nu} \},
\end{equation}
where $\delta / \delta g^{\mu\nu}$ represents the variation with respect to the metric, while $G_{\mu\nu}$ is some  fermionic tensor. Thus, the fact that the action is $\mathcal{Q}$-exact implies the same for the energy-momentum tensor. The above equation indeed reflects \eqref{susy algebra topological}, where all components of the energy-momentum tensor and hence momentum are $\mathcal{Q}$-exact, whence correlation functions of the $\cal Q$-cohomology of operators in the theory are independent of spacetime coordinates, i.e., they are topological or metric-independent:
\begin{equation}
    \frac{\delta}{\delta g^{\mu\nu}}\langle \Pi_i \mathcal{O}_i\rangle=\frac{\delta}{\delta g^{\mu\nu}}\int_{\mathcal{M}}D\phi\; \Pi_i\mathcal{O}_i e^{-S}=\langle T_{\mu\nu}\Pi_i \mathcal{O}_i\rangle = 0.
\end{equation}
Here, $D\phi$ represents the integration measure over all field configurations $\phi$, and $\mathcal{O}_i$ are operators in the $\mathcal{Q}$-cohomology. To arrive at this, we used the fact that the $\mathcal{O}_i$'s are $\mathcal{Q}$-closed, and that $\langle \{{\cal Q}, \cdots \} \rangle = 0$ by supersymmetry.

\bigskip\noindent\textit{The BPS Equations}
\vspace*{0.5em}\\
Another consequence of \eqref{Q exact action} is that the theory is independent of the coupling constant $e$ - by a similar calculation to \eqref{enery-momentum Q exact}, we find that $\delta S / \delta e \sim 0$ in $\cal Q$-cohomology, so $\frac{\delta}{\delta e}\langle \Pi_i \mathcal{O}_i\rangle= 0$.
%since a variation with respect to $e$ is equivalent to computing an expectation value of a $\mathcal{Q}$-exact correlation function, which in turn is trivial in $\mathcal{Q}$-cohomology.
We can thus go to the limit of weak coupling, where exact computations of correlation functions can be performed via a semi-classical expansion around the BPS field configurations that the path integral localises on. From $\delta \psi=0$ (where $\psi$ represents a generic fermionic field) which minimises the action in \ref{Q exact action}, the corresponding BPS equations are:
\begin{subequations}\label{4d BPS}
   \begin{empheq}[box=\widefbox]{align}
     F_{\bar{w}\bar{z}}+i [B_{\bar{z}}, B_{\bar{w}}]&=0\label{4d BPS a}\\
  F_{wz}-i[B_{z}, B_{w}]&=0\label{4d BPS b}\\
    (F_{z\bar{z}} -i[B_{z}, B_{\bar{z}}])g^{z\bar{z}}-i[C, C^{\dagger}]&\notag\\
   +(F_{w\bar{w}}+i[B_{\bar{w}}, B_{w}])g^{w\bar{w}} &=0 \label{4d BPS c}\\
  g^{z\bar{z}} D_{\bar{z}}B_{z} -g^{w\bar{w}} D_{w}B_{\bar{w}}&=0\label{4d BPS d}\\
   D_{\bar{w}}B_{z} + D_{z}B_{\bar{w}}&=0 \label{4d BPS e}\\
    g^{w\bar{w}}D_{\bar{w}}B_{w}+ g^{z\bar{z}}D_{z}B_{\bar{z}}&=0 \label{4d BPS f}\\
    D_{\bar{z}}B_{w}-D_{w}B_{\bar{z}}&=0 \label{4d BPS g}\\
    [C, B_{\bar{z}}]=[C, B_{z}]&=0\label{4d BPS h}\\
    [C, B_{w}]=[C, B_{\bar{w}}]&=0\label{4d BPS i}\\
    D_{\mu}C&=0 \label{4d BPS j}
\end{empheq} 
\end{subequations}
We first note from \eqref{4d BPS j} that non-trivial solutions to $D_{\mu}C=0$ correspond to reducible gauge connections, in turn leading to an ill-behaved moduli space $\mathcal{M}$ of solutions to the BPS equations \ref{4d BPS}.  To avoid complications in $\mathcal{M}$, we want irreducible gauge connections, so we shall set (the zero modes of) $C=0$. \emph{Novel} $\mathcal{M}$ will then consist of configurations of $A_{\mu}, B_{\mu}$ (zero modes) that satisfy \eqref{4d BPS}. 
%$A_{\mu}, B_{\bar{w}}, B_{w}, B_{z}, B_{\bar{z}}$ satisfying \eqref{4d BPS}. 

\bigskip\noindent\textit{Ghost Number of the Fields}
\vspace*{0.5em}\\
Let us now examine the ghost number symmetry of the fields. 
With the conventional definition of $\mathcal{Q}$ having a ghost-number of 1, a consistent assignment for the ghost-numbers $gh$  of the fields will be as follows:
\begin{equation}\label{ghost numbers}
    \begin{aligned}
    gh &=2 \quad &: \quad &C\,,\\
    gh &=1\quad &: \quad & \lambda_{1z}, \,\lambda_{1w},\, \,\lambda_{3\bar{w}},\,\lambda_{3\bar{z}},\,\bar{\lambda}_{2\bar{z}}, \, \bar{\lambda}_{2w}\, \bar{\lambda}_{4z}, \,\bar{\lambda}_{4\bar{w}},\\
    gh & =0 \quad &: \quad &A_{\mu}, B_{\mu}\\ %\,B_{\bar{w}},\,B_{w},\,B_{z},\, B_{\bar{z}},\, \\
     gh & =-1 \quad &: \quad & \lambda_{2}, \,\lambda_{4},\,\bar{\lambda}_{1},\,\bar{\lambda}_{3},\,\lambda_{2\bar{w}z},\,\lambda_{4w\bar{z}},\, \bar{\lambda}_{1\bar{w}\bar{z}}, \,\bar{\lambda}_{3wz},\\
      gh &=-2 \quad &: \quad &C^{\dagger}\,.\\
    \end{aligned}
\end{equation}
It is worth mentioning that the two scalar supercharges $\bar{\mathcal{Q}}_1$ and $\bar{\mathcal{Q}}_3$ have the same ghost-number of $gh=1$. This is similar to the situation in the twist studied in \cite{kapustin2006electric} (the Geometric-Langlands twist), where the topological supercharge is also written as a linear combination of the two scalar supercharges. The index of such theories is generally non-zero, and we can consider non-trivial correlation functions in the theory beyond the partition function. The virtual dimension of $\mathcal{M}$ is a topological invariant, and it is given by the index $k$ of the theory. It is equal to the difference between the number of zero-modes of 1-form and 0- and 2-form fermions.
In the case where the 0- and 2-form fermions have no zero-modes, the virtual dimension will be equal to the actual dimension of $\mathcal{M}$.
%The supersymmetries that are preserved for a generic curved $M_4$ are those that are scalars on both $\Sigma_{1}$ and $C$.
 
In contrast, the twist studied in \cite{vafa1994strong} (the Vafa-Witten twist) has two scalar supercharges of \emph{opposite} ghost-numbers $gh=\pm1$. A theory like this belongs to class of TQFT's called ``balanced topological field theories" \cite{dijkgraaf1997balanced} where the index and virtual dimension of the moduli space that the theory localizes on is zero. The only non-trivial observable of VW theory is thus the partition function (when there are no 0- and 2-form fermion zero modes).

\bigskip\noindent\textit{Topological Correlation Functions}
\vspace*{0.5em}\\
Non-zero correlation functions in $\mathcal{Q}$-cohomology will have to satisfy an anomaly cancellation condition, where the total ghost number of the operators in a correlation function must be equal to the index of the theory. 
Topological observables of the theory consists of those which are in $\mathcal{Q}$-cohomology i.e. $\mathcal{Q}$-closed operators that are not $\mathcal{Q}$-exact. As per the standard prescription in topological gauge theories, the gauge-invariant local operator in $\mathcal{Q}$-cohomology is given by\footnote{$\cal Q$-cohomology operators are constructed out of the zero-modes of their constituent fields. However, recall that we have set the zero modes of $C$ to 0. Therefore, to construct the gauge-invariant local operator in \eqref{topological local operator}, one has to define it in terms of $\langle C \rangle$, which will then allow us to effectively express it in terms of the zero modes of other fields. This procedure is sketched out in \cite{witten1988topological}. }
\begin{equation}\label{topological local operator}
    \mathcal{O}^{(0)} = \text{Tr}\,\langle C \rangle^{N}, \quad \text{for appropriate} \quad N\in\mathbb{Z}^+.
\end{equation}
Starting with $\mathcal{O}^{(0)}$ one can obtain higher order non-local operators $\mathcal{O}^{(m)}$, $m=1,2,3,4$ by following the descent procedure as in \cite{witten1988topological}, where $\mathcal{O}^{(1)}$ is a 1-form operator integrated over a one-cycle on $M_4$ and so on. The general form of a topological correlation function is
\begin{equation}\label{4d topo corr function}
    \langle \Pi_i\mathcal{O}^{(m)}_i \rangle_{4d}\big(\tau, G\big)= \int_{\mathcal{M}}D\phi\; \Pi_i\mathcal{O}^{(m)}_i e^{-S},
\end{equation}
where $D\phi$ represents the integration measure over all field configurations, $\sum_i gh_i=k$, and
\begin{equation}
    {\mathcal{O}}^{(m)}_i \cong H^{*}_{\text{DR}}(\mathcal{M}, \mathbb R) \otimes H_{m}(M_4, \mathbb R).
\end{equation}
The non-trivial dependence on $\tau$ is due to the topological term in \ref{Q exact action}, which contributes an overall $\tau$-dependent multiplicative factor to correlation functions. We also choose to indicate explicitly the dependence of correlation functions on the gauge group $G$.

\bigskip\noindent\textit{A Hybrid Between VW and KW Theory}
\vspace*{0.5em} \\
Recall that in \eqref{first top susy charge}, $s=b/a\neq 0, \infty$, where $\mathcal{Q}$ is a linear combination of scalar supercharges. It is interesting to note that this choice of $\mathcal{Q}$-cohomology results in a topological theory which effective twist is a hybrid of the Vafa-Witten twist \cite{vafa1994strong} and the Geometric Langlands (GL) twist \cite{kapustin2006electric}. Like the VW equations, the BPS equations \eqref{4d BPS} contain commutators, yet like the Kapustin-Witten (KW) equations, the bosonic fields consist only of scalars and one-forms with no two-forms. Because of the commutators, we will not be able to write the BPS equations in terms of a 4d complexified gauge connection, as is the case with the VW equations. In contrast, one is able to do so with the KW equations.

\bigskip\noindent\textit{About the Parameter $s$}
\vspace*{0.5em} \\
This now brings us to the complex parameter $s$, which we have earlier set to $s=1$. Note that $s$ would enter the KW equations when they are expressed in terms of a 4d complexified gauge connection. Consequently, upon dimensional reduction along $\Sigma_1$, we will have a family of effective theories on $\Sigma_2$ parameterized by $s$. Thus, the value of $s$ would be relevant to the 2d theory if the 4d BPS equations can be expressed in terms of a 4d complexified gauge connection.  

As pointed out in the paragraph before last, we cannot express our 4d BPS equations in terms of a 4d complexified gauge connection. As such, $s$ will play no role in the effective 2d theory on $\Sigma_2$, and we will indeed obtain a unique theory in 2d, as we shall show in the next section. In turn, by the topological invariance of our theory along all of $M_4 = \Sigma_1 \times \Sigma_2$, it would mean that the value of $s$ will not be relevant in the 4d theory either. This justifies our earlier  setting of $s =1$ for convenience.

%While not as apparent in 4d, the value of that this parameter takes will directly impact the target space of the sigma-model of the theory with the GL twist upon shrinking $M_4$ to two dimensions. Thus, we see that the relevance of $s$ comes when we are able to have a complexified gauge connection in the theory. Since this is not the case at hand, the theory remains the same for $s\neq 0, \infty$, and it is sufficient to examine the case where $s=1$. The implications of the resemblance of \eqref{4d BPS} to the VW equations will be more apparent when we shrink $\Sigma_{1}$ to obtain an effective sigma-model on $\Sigma_2$, which will be explored in the next section. 

%Also, this theory generically has a non-zero index, similar to the Geometric Langlands twist where there are non-trivial correlation functions other than the partition function. 

%This concludes the discussion of $\mathcal{Q}=\bar{\mathcal{Q}}_1+\bar{\mathcal{Q}}_3$ cohomology on the four dimensional theory. We wish to establish a duality by exploiting the fact that this theory is independent of metric deformations, where we shrink $\Sigma_{1}$ to obtain an effective 2d topological sigma-model on $C$.

\section{An \texorpdfstring{$\mathcal{N}=(4,4)$}{N} Topological \texorpdfstring{$A$}{A}-model on $\Sigma_2$ and a 4d-2d Correspondence}\label{section: topological A-model 2d}

In this section, we will first perform a dimensional reduction of the 4d theory along $\Sigma_1$ down to 2d, where the method employed will be that in \cite{bershadsky1995topological}.
We will show that an $\mathcal{N}=(4,4)$ topological $A$-model in complex structure $I$ is obtained on $\Sigma_2$. Then, by the topological invariance of the 4d theory along $\Sigma_1$, we will arrive at a 4d-2d correspondence. 

\bigskip\noindent\textit{Dimensional Reduction Along $\Sigma_1$}
\vspace*{0.5em} \\
To perform dimensional reduction along $\Sigma_1$, we write the metric as 
\begin{equation}
    g_M = \text{diag}\big( \varepsilon g_{\Sigma_{1}},  g_{\Sigma_2} \big),
\end{equation}
where $\varepsilon$ is a small parameter to deform $g_{\Sigma_{1}}$. Deforming the metric inevitably affects the terms in the action, since they involve contraction of indices by the metric tensor. With the introduction of the $\varepsilon$ parameter, the determinant changes by $\sqrt{g}\to \varepsilon\sqrt{g}$. Thus, terms that survive after taking the limit $\varepsilon\to0$ have one contraction of indices $z,\bar{z}$ on $\Sigma_{1}$ by the metric tensor $g^{z\bar{z}}$, giving a factor of $\varepsilon^{-1}$ to cancel a factor of $\varepsilon$ from the determinant.

Terms in the action with more than one contraction of indices $z,\bar{z}$ have higher orders of $\varepsilon^{-1}$, which blows up in the limit $\varepsilon\to0$.\footnote{There are no fermionic terms with more than one contraction of $z,\bar{z}$ by the metric tensor $g^{z\bar{z}}$.} 
We are thus forced to set these terms to zero to keep the action finite. This is also equivalent to selecting the 4d BPS equations with indices belonging only to $\Sigma_{1}$.
We then have the finiteness condition from \eqref{4d BPS c} and \eqref{4d BPS d} as
\begin{subequations}\label{finiteness condition topological}
   \begin{empheq}[box=\widefbox]{align}
    F_{z\bar{z}}-i[B_{z},B_{\bar{z}}]&=0\\
    D_{\bar{z}}B_{z}&=0
\end{empheq} 
\end{subequations}
These are Hitchin's equations, and the space of solutions of $(A_{\Sigma_{1}}, B_{\Sigma_{1}})$ modulo gauge transformations then span Hitchin's moduli space $\mathcal{M}^G_H(\Sigma_{1})$ for a connection $A_{\Sigma_{1}}$ on a principal $G$-bundle $P$ over $\Sigma_{1}$, and a section $B_{\Sigma_{1}}\in\Omega^1(\Sigma_{1})$. The above equations leave the $(w, \bar{w})$ dependence of $A_{\Sigma_{1}}$ and $B_{\Sigma_{1}}$ arbitrary, and thus the fields $(A_{\Sigma_{1}}, B_{\Sigma_{1}})$ define a map $\Phi: \Sigma_2 \to \mathcal{M}^G_H(\Sigma_{1})$. This moduli space is well understood to be hyperK\"{a}hler, with three complex structures $I, J, K$ obeying quarternionic relations. Since $\mathcal{M}^G_H(\Sigma_{1})$ has dimension $\text{dim}_{\mathbb{C}}\mathcal{M}^G_H=2\text{dim}_{G}(g-1)$, we will take $\Sigma_{1}$ to have a genus $g\geq 2$ for simplicity. 

\bigskip\noindent\textit{The Surviving 4d Action}
\vspace*{0.5em}\\
As mentioned, we keep only terms in the action that have a single contraction with $g^{z \bar z}$. The surviving 4d action (excluding the topological term) is then 
\begin{equation}\label{eff 2d action}
    \begin{aligned}
    S_{\text{surv}} &= \frac{1}{e^2}\int_{M_4} d^2z d^2w \sqrt{g}\,\text{Tr}\bigg[\frac{1}{2}(F_{zw}F^{zw}+F_{\bar{z}w}F^{\bar{z}w}+F_{z\bar{w}}F^{z\bar{w}}+F_{\bar{z}\bar{w}}F^{\bar{z}\bar{w}})+g^{z\bar{z}}g^{w\bar{w}}(D_{w}B_{z}D_{\bar{w}}B_{\bar{z}}\\
    &+D_{\bar{w}}B_{z}D_{w}B_{\bar{z}}
    -i\bar{\lambda}_{1\bar{w}\bar{z}}D_{w}\lambda_{1z}-i\bar{\lambda}_{2\bar{z}}D_{w}\lambda_{2\bar{w}z}-i\bar{\lambda}_{3wz}D_{\bar{w}}\lambda_{3\bar{z}}-i\bar{\lambda}_{4z}D_{\bar{w}}\lambda_{4w\bar{z}})+\dots\bigg]
    \end{aligned}
\end{equation}
where only the kinetic energy terms are shown, and the ``$\dots$" represent interaction terms that survive in the limit $\varepsilon\to0$. Note also that terms with derivatives on $\Sigma_{1}$ (i.e. $D_{z}$ or $D_{\bar{z}}$) are also  omitted in \eqref{eff 2d action}, as we are ultimately reducing the 4d theory along $\Sigma_1$. 

%One can see that this is in the form of an $\mathcal{N}=(4,4)$ sigma model, as expected from the sigma model having a  hyperK\"{a}hler target space \cite{alvarez1981geometrical}. The expectation that an $\mathcal{N}=(4,4)$ supersymmetry is obtained can also be seen by inspecting the supercharges, where the supercharges that survive upon shrinking of $\Sigma_{1}$ are those that are scalar on $\Sigma_{1}$ and there are exactly 8 of them of opposite chirality (see Table \ref{chargetable}).

\subsection{A Topological Sigma Model in Complex Structure $I$ and Gromov-Witten Invariants}

%\bigskip\noindent\textit{A Topological Sigma Model in Complex Structure $I$}
%\vspace*{0.5em}\\
As mentioned, the fields $(A_{\Sigma_{1}}, B_{\Sigma_{1}})$ define a map $\Phi: \Sigma_2 \to \mathcal{M}^G_H(\Sigma_{1})$. This means that the 4d theory should reduce to a 2d sigma model on $\Sigma_2$ with  hyperK\"{a}hler target $\mathcal{M}^G_H(\Sigma_{1})$. Let us now determine what this 2d sigma model is.

%To this end, we first note that it is \emph{not} possible to express \eqref{4d BPS} in terms of a flat complexified connection, unlike the KW equations. In the theory with a GL twist that leads to the KW equations, the scalar supercharge (as a combination of two different scalar supercharges) also has a parameter $t\in\mathbb{C}$, where in the effective sigma-model after shrinking to two dimensions, the value of $t$ determines the complex structure of the hyperK\"{a}hler target. This is because the finiteness conditions resulting from the KW equations can be expressed in terms of a flat complexified connection on the shrunken Riemann surface. On the other hand, this is not possible for the VW equations, where the finiteness conditions are always that of a Hitchin pair. The same goes for the BPS equations of this theory. Thus, for $s\neq 0, \infty$, the behaviour of the effective sigma-model of our theory under this choice of $\mathcal{Q}$-cohomology is expected to be like that in VW theory, where we obtain a sigma-model in complex structure $I$.

%Note also, that the parameter $s$ does not appear in \eqref{finiteness condition topological}. 

To this end, first, recall that as pointed out at the end of $\S$\ref{section: 4d topological theory}, the 2d sigma model is unique, regardless of the value of the parameter $s$. In fact, we ought to have a sigma model in complex structure $I$ of $\mathcal{M}^G_H(\Sigma_{1})$. 

The fact that we end up in complex structure $I$ can be verified by examining the topological term in the 4d action, where one can show that the surviving component of the gauge field strength is of the form $F\,\propto\, \omega_{I}$, the K\"{a}hler form of Hitchin's moduli space in complex structure $I$ \cite{ong2022vafa}. In this complex structure, Hitchin's moduli space is equivalent to $\mathcal{M}_{\text{Higgs}}^{G}(\Sigma_{1})$, the moduli space of stable Higgs $G$-bundles on $\Sigma_{1}$.

Second, we can assign a basis $(\alpha_{i\bar{z}}, \alpha_{\bar{i}z})$ and $(\beta_{iz}, \beta_{\bar{i}\bar{z}})$ to the base and fiber of $\mathcal{M}^G_H(\Sigma_{1})$, respectively, where $i, \bar{i}=1,2,\dots, \frac{1}{2}\text{dim}_{\mathbb{C}}\big(\mathcal{M}^G_H(\Sigma_{1}) \big)$. The fermion fields can be expressed in the linear combinations
\begin{equation}
\begin{aligned}\label{2d fermion linear combi}
    \bar{\lambda}_{2\bar{z}}&=\psi_{2}^{\bar{i}}\beta_{\bar{i}\bar{z}}, \quad &\lambda_{1z}&=\psi^{\bar{i}}_{1}\alpha_{\bar{i}z}, \\
    \lambda_{2\bar{w}z}&=\rho_{2\bar{w}}^{i}\beta_{iz} \quad & \bar{\lambda}_{1\bar{w}\bar{z}}&=\rho^{i}_{1\bar{w}}\alpha_{i\bar{z}},\\
    \bar{\lambda}_{4z}&=\psi_{2}^{i}\beta_{iz}, \quad &\lambda_{3\bar{z}}&=\psi^{i}_{1}\alpha_{i\bar{z}}, \\
    \lambda_{4w\bar{z}}&=\rho_{2w}^{\bar{i}}\beta_{\bar{i}\bar{z}} \quad & \bar{\lambda}_{3wz}&=\rho^{\bar{i}}_{1w}\alpha_{\bar{i}z},
\end{aligned}
\end{equation}
where $\psi^{i}_{1}, \psi^{\bar{i}}_{1}, \psi^{i}_{2}, \psi^{\bar{i}}_{2}, \rho^{i}_{1\bar{w}}, \rho^{\bar{i}}_{1w}, \rho^{i}_{2\bar{w}}, \rho^{\bar{i}}_{2w}$ are two-dimensional fermionic fields on $\Sigma_2$.

Third, let us determine the BPS equations of the effective 2d theory which can be obtained via dimensional reduction of the 4d BPS equations. We need only consider \eqref{4d BPS a} and \eqref{4d BPS e}, since they are the only ones (in mixed coordinates) that relate to the terms which appear in \eqref{eff 2d action}, whence they would lead to the relevant 2d BPS equations.\footnote{The other 4d BPS equations which are relevant are \eqref{4d BPS b} and \eqref{4d BPS g}, but they are just complex conjugates of \eqref{4d BPS a} and \eqref{4d BPS g}, so we shall not need to consider them.} We can take $F_{\bar{w}\bar{z}}=\partial_{\bar{w}}A_{\bar{z}}$ and $D_{\bar{w}}B_z=\partial_{\bar{w}}B_z$ etc, since $A_{\bar{w}}$ does not have derivatives on $\Sigma_2$ and is thus a non-dynamical field that can be integrated out via its equation of motion (which expresses it as a combination of fermionic fields). %The same goes for $A_{w}$. 

With the following correspondences for the bosonic fields,
\begin{equation}\label{field corres}
    \begin{aligned}
    A_{\bar{z}} 	&\leftrightarrow X^i, &\qquad B_{z} 	&\leftrightarrow Y^i,\\
    A_{z}&\leftrightarrow X^{\bar{i}}, &\qquad  B_{\bar{z}}&\leftrightarrow Y^{\bar{i}},
    \end{aligned}
\end{equation}
one can compute the 2d BPS equations to be 
\begin{equation} \label{2d BPS equations}
    \boxed{\partial_{\bar{w}}X^{i}=\partial_{\bar{w}}Y^{i}=0}
\end{equation}

Last but not least, the supersymmetry transformations of the fields in the sigma-model can be written as
\begin{equation}\label{2d bps topo}
    \begin{aligned}
   \delta X^i &= \psi^{i}_{1}, &\quad \delta Y^i &= \psi^{i}_{2},\\
   \delta X^{\bar{i}} &= \psi^{\bar{i}}_{1}, &\quad \delta Y^{\bar{i}} &= \psi^{\bar{i}}_{2},\\
   \delta \psi^{i}_{1}&=0, &\quad \delta \psi^{i}_{2}&=0,\\
   \delta \psi^{\bar{i}}_{1}&= 0, &\quad \delta \psi^{\bar{i}}_{2}&= 0,\\
   \delta \rho^{i}_{1\bar{w}}&= \partial_{\bar{w}}X^i+\dots, &\quad \delta \rho^{i}_{2\bar{w}}&= \partial_{\bar{w}}Y^i+\dots,\\
   \delta \rho^{\bar{i}}_{1w}&= \partial_{w}X^{\bar{i}}+\dots, &\quad \delta \rho^{\bar{i}}_{2w}&= \partial_{w}Y^{\bar{i}}+\dots,
    \end{aligned}
\end{equation}
where ``$\dots$'' represent terms involving the curvature of the target space $\mathcal{M}_{\text{Higgs}}^{G}(\Sigma_{1})$. These are exactly the transformations of an $\mathcal{N}=(4,4)$ topological sigma model, and together with \ref{2d BPS equations}, we conclude that it is an $A$-model where %$\mathcal{Q}$ is interpreted as a de Rham operator on $\mathcal{M}_{\text{Higgs}}^{G}(\Sigma_{1})$, and% 
the path integral localizes to an integral over the moduli space of holomorphic maps
\begin{equation} \label{Mmaps}
    \boxed{\mathcal{M}_{\text{maps}} = \{ \Phi(X^i, Y^i):\Sigma_2\to\mathcal{M}^G_{\text{Higgs}}(\Sigma_{1}) \;|\; \partial_{\bar{w}}X^{i}=\partial_{\bar{w}}Y^{i}=0 \}}
\end{equation}

In short, we obtain an $\mathcal{N}=(4,4)$ topological $A$-model in complex structure $I$, with action
\begin{equation}\label{2dfinalaction2}
    \boxed{S_1=\frac{1}{e^2}\int_{\Sigma_2}|dw^2| g_{i\bar{j}}\bigg(\partial_{w}X^{\bar{i}}\partial_{\bar{w}}X^{j}+\partial_{w}X^{i}\partial_{\bar{w}}X^{\bar{j}}+\partial_{w}Y^{\bar{i}}\partial_{\bar{w}}Y^{j}+\partial_{w}Y^{i}\partial_{\bar{w}}Y^{\bar{j}} \bigg)+i\tau\int_{\Sigma_2}\Phi^{*}(\omega_I)+ \dots}
\end{equation}
where the second term in $\tau$ is the topological worldsheet instanton term, and ``\dots" represent fermionic terms. %and the integration measure and the last term is a topological worldsheet instanton term in \eqref{2dfinalaction2}, a reflection that the effective topological $A$-model is in complex structure $I$. 

\bigskip\noindent\textit{Gromov-Witten Invariants}
\vspace*{0.5em}\\
Since we have an $A$-model, the topological correlation functions of the theory
\begin{equation}\label{2d topo A model correlation function}
    \langle \Pi_i\tilde{\mathcal{O}}^{(p)}_i \rangle_{2d} \big(\tau, \mathcal{M}^G_{\text{Higgs}}(\Sigma_{1})\big)= \int_{\mathcal{M}_{\text{maps}}}D\tilde{\phi}\; \Pi_i\tilde{\mathcal{O}}^{(p)}_i e^{-S_1}
\end{equation}
correspond to Gromov-Witten (GW) invariants of $\mathcal{M}^G_{\text{Higgs}}(\Sigma_{1})$, where $\tilde{\phi}$ represents generic fields in the effective sigma-model, and
\begin{equation}
\tilde{\mathcal{O}}^{(p)}_i \cong H^{*}_{\text{DR}}({\mathcal M}_{\text{maps}}, \mathbb R) \otimes H_{p}(\Sigma_2, \mathbb R), \qquad  p = 0,1,2.
\end{equation}
 The degree of $\tilde{\mathcal{O}}^{(p)}_i$ as a differential form in $\mathcal{M}_{\text{maps}}$ corresponds to its total ghost number, which is inherited from the respective fields in 4d.

Since the index $k$ of the 4d theory is a topological invariant, the $A$-model will inherit the same index, where the anomaly cancellation condition $\sum_i gh_i=k$ applies as well.
%Operators $\mathcal{O}$ that are inserted in \eqref{4d topo corr function} thus have their counterparts in the fields of the 2d $A$-model according to \eqref{2d fermion linear combi} and \eqref{field corres}. Recall that fields in 4d that do not survive the reduction (for example, $A_{C}$) are integrated out via their equations of motion. Insertions of such fields in $\mathcal{O}$ in \eqref{4d topo corr function} will then generally correspond to insertions of fermionic fields in \eqref{2d fermion linear combi} and curvature terms of $\mathcal{M}^G_{\text{Higgs}}(\Sigma_{1})$ into \eqref{2d topo A model correlation function}.
%Similar to the earlier analysis on VW theory, due to the fact that the BPS equations of the 4d gauge theory reduce directly to that of 2d sigma-model, the moduli space $\mathcal{M}$ to $\mathcal{M}_{\text{maps}}$, and  $\text{dim}\mathcal{M}_{\text{maps}}=k$. This is due to the fact that 
Note also that the K\"{a}hler form $\omega_I$ is not $\mathcal{Q}$-exact, so \eqref{2d topo A model correlation function} will have  non-trivial contributions from the topological worldsheet instanton term, and therefore a dependence on the complexified coupling parameter $\tau$. 
%This is consistent with the fact that there is also a $\tau$-dependence in the 4d correlation functions, from which the $A$-model correlation functions actually originate from. 
This observation will be important shortly. 

\subsection{A 4d-2d Correspondence between Topological Invariants of $M_4$ and Gromov-Witten Invariants of Space of Higgs Bundles}

We are now ready to establish a 4d-2d correspondence. With respect to $\mathcal{Q}=\bar{\mathcal{Q}}_1+\bar{\mathcal{Q}}_3$ cohomology, metric independence of the 4d theory along $\Sigma_1$ allows us to establish a correspondence between the 4d topological invariants of $M_4$ and 2d GW invariants which define a quantum cohomology ring of $\mathcal{M}^G_{\text{Higgs}}(\Sigma_{1})$:
\begin{equation}\label{4d-2d corr top}
   \boxed{ \langle \Pi_i\mathcal{O}^{(m)}_i \rangle_{4d}\big(\tau, G\big) = \langle \Pi_i\tilde{\mathcal{O}}^{(p)}_i \rangle_{2d} \big(\tau, \mathcal{M}^G_{\text{Higgs}}(\Sigma_{1})\big)}
\end{equation}

The equality sign in \eqref{4d-2d corr top} reflects the fact that the 4d operators $\mathcal{O}^{(m)}_i$ will have corresponding counterparts $\tilde{\mathcal{O}}^{(p)}_i$ in the 2d sigma-model. Inevitably, there will be some fields that make up $\mathcal{O}^{(m)}_i$ which will become auxiliary in the reduction along $\Sigma_1$. Nonetheless, they can, via equations of motion, be expressed in terms of the physical fields of the sigma-model that make up $\tilde{\mathcal{O}}^{(p)}_i$ (see, for example, \cite{bershadsky1995topological}). Last but not least, `$p$' will be given by the number of directions along $\Sigma_2$ of the $m$-cycle in $M_4$.

\section{A Topological-Holomorphic Theory on $M_4$}\label{section: 4d top-hol theory}

In this section, we consider a different $\mathcal{Q}'$-cohomology of the topological-holomorphic twist, where
\begin{equation} \label{Q'}
    \mathcal{Q}'=u\bar{\mathcal{Q}}_3 + v \mathcal{Q}_4
\end{equation}
and $u,v\in \mathbb{C}$. Here, we consider the case where $t=v/u\neq 0, \infty$.
Similar to before, we first single out the supersymmetry algebra involving the anti-commutators of $\bar{\mathcal{Q}}_3$ and $\mathcal{Q}_4$. From the (twisted) supersymmetry relations \eqref{susy algebra general}, we find that
\begin{equation}
\begin{aligned}\label{susy algebra top-hol}
    \{\bar{\mathcal{Q}}_3, \mathcal{Q}_{3\bar{w}} \} &\propto P_{\bar{w}},\\
    \{\bar{\mathcal{Q}}_3, \mathcal{Q}_{3\bar{z}}\} &\propto P_{\bar{z}},\\
    \{\mathcal{Q}_4, \bar{\mathcal{Q}}_{4z}\} &\propto P_{z},\\
    \{\mathcal{Q}_4, \bar{\mathcal{Q}}_{4\bar{w}}\} &\propto P_{\bar{w}}.\\
\end{aligned}
\end{equation}
Noting that $\{\mathcal{Q}_i, \mathcal{Q}_j \}=\{\bar{\mathcal{Q}}_i, \bar{\mathcal{Q}_j} \}=0$,  we compute that
\begin{equation} \label{top charge hol in w}
\begin{aligned}
       \{\mathcal{Q}', \mathcal{Q}_{3\bar{w}} \} \propto P_{\bar{w}},\\
    \{\mathcal{Q}', \mathcal{Q}_{3\bar{z}} \} \propto P_{\bar{z}},\\
    \{\mathcal{Q}', \bar{\mathcal{Q}}_{4\bar{w}} \} \propto P_{\bar{w}},\\
    \{\mathcal{Q}', \bar{\mathcal{Q}}_{4z} \} \propto P_{z}.\\
\end{aligned}
\end{equation}
Since $P_\mu \propto \del_\mu$, the above implies that there is now a dependence of the theory on the $w$-coordinate in $\mathcal{Q}'$-cohomology. In other words, the theory is topological along $\Sigma_{1}$ but holomorphic along 
$\Sigma_2$.

The transformations of the fields under $\mathcal{Q}'$ are
\begin{equation} \label{4d susy transformations top-hol}
    \begin{aligned}
        \delta' A_{z}  &= -i\epsilon_{4} \bar{\lambda}_{4z}\;, & \quad\quad \delta' A_{\bar{z}}&= -i\bar{\epsilon}_{3}\lambda_{3\bar{z}}\;,\\
        \delta' A_w &= 0\;,& \quad\quad\delta' A_{\bar{w}}&= -i\epsilon_{4} \bar{\lambda}_{4\bar{w}}-i\bar{\epsilon}_{3}\lambda_{3\bar{w}}\;,\\
        \delta' B_{\bar{w}} &= 0\;, & \quad\quad \delta' B_{w}&=-\epsilon_{4}\lambda_{1w}-\bar{\epsilon}_{3}\bar{\lambda}_{2w}\;,\\
        \delta' C &=0\;, & \quad\quad  \delta' C^{\dagger}&=i(\epsilon_{4}\lambda_{2}-\bar{\epsilon}_{3}\bar{\lambda}_{1})\;,\\
        \delta' B_{z} &=\bar{\epsilon}_{3} \bar{\lambda}_{4z}\;, & \quad\quad\delta' B_{\bar{z}}&=\epsilon_{4}\lambda_{3\bar{z}}\;,\\
        \delta' \lambda_{1z}  &= -\bar{\epsilon}_{3}D_{z}C+\epsilon_{4} [B_{z},C] \;, & \quad\quad
        \delta' \lambda_{1w}  &= -\bar{\epsilon}_{3} D_{w}C\;, \\
        \delta' \lambda_{2}  &= -i\bar{\epsilon}_{3}g^{w\bar{w}}D_{w}B_{\bar{w}}\;, & \quad\quad \delta' \lambda_{2\bar{w}z}&=-i\bar{\epsilon}_{3}D_{z}B_{\bar{w}}-i\epsilon_{4} [B_{\bar{w}}, B_{z}]\;,\\
        \delta' \lambda_{3\bar{w}}  &= \epsilon_{4}[B_{\bar{w}}, C]& \quad\quad\delta' \lambda_{3\bar{z}} &= 0\;,\\    
        \delta' \lambda_{4}  &=\big(\epsilon_{4} F_{z\bar{z}}-i\epsilon_{4}[B_{z}, B_{\bar{z}}]-i\bar{\epsilon}_{3} D_{z}B_{\bar{z}}\big)g^{z\bar{z}} \; & \quad\quad\delta' \lambda_{4w\bar{z}}&=\epsilon_{4} F_{w\bar{z}}+i\bar{\epsilon}_{3}D_{w}B_{\bar{z}}\;,\\
        &-\epsilon_{4}g^{w\bar{w}}\big(F_{w\bar{w}}+i[B_{\bar{w}}, B_{w}]\big)+i\epsilon_{4}[C, C^{\dagger}]\;,\\
        \delta'\bar{\lambda}_{1}  &= -i\epsilon_{4} D_{w}B_{\bar{w}}\;, & \quad\quad\delta' \bar{\lambda}_{1\bar{w}\bar{z}}&= -i\epsilon_{4} D_{\bar{z}}B_{\bar{w}}+i\bar{\epsilon}_{3}[B_{\bar{z}}, B_{\bar{w}}]\;,\\
        \delta'\bar{\lambda}_{2\bar{z}}  &=\epsilon_{4} D_{\bar{z}}C+\bar{\epsilon}_{3}[C, B_{\bar{z}}]\;, & \quad\quad
        \delta'\bar{\lambda}_{2w} &= \epsilon_{4} D_{w}C\;,\\
        \delta'\bar{\lambda}_{3} &=-\big(\bar{\epsilon}_{3} F_{z\bar{z}}-i\bar{\epsilon}_{3}[B_{z}, B_{\bar{z}}]+i\epsilon_{4} D_{\bar{z}}B_{z})g^{z\bar{z}}\; & \quad\quad\delta' \bar{\lambda}_{3wz}&= -\bar{\epsilon}_{3}F_{wz}-i\epsilon_{4} D_{w}B_{z}\;,\\
        &-\bar{\epsilon}_{3}g^{w\bar{w}}\big(F_{w\bar{w}}+i[B_{\bar{w}}, B_{w}]\big)+i\bar{\epsilon}_{3}[C, C^{\dagger}]\;,\\
        \delta' \bar{\lambda}_{4\bar{w}} &=-\bar{\epsilon}_{3}[B_{\bar{w}}, C]\;, & \quad\quad\delta' \bar{\lambda}_{4z}&=0\;.
    \end{aligned}
\end{equation}
%The variation under $\mathcal{Q}'$ can be obtained by setting $\bar{\epsilon}_3$ to be proportional to $\epsilon_4$. 
Notice that $\bar{\epsilon}_3$ and $\epsilon_4$ are supposed to be proportional to $u$ and $v$ in \ref{Q'}, respectively. This means that we can let  $\bar{\epsilon}_3=it^{-1}\epsilon_4$ henceforth.

Note that the variation $\delta'$ is nilpotent on all fields except for $\lambda_4$ and $\bar{\lambda}_3$. We can remedy this by introducing the linear combinations
\begin{equation}
    \eta = \lambda_{4}+ it\bar{\lambda}_{3} , \qquad \gamma = \lambda_{4}  -it\bar{\lambda}_{3},
\end{equation}
where $\delta'^2 \eta=0$, although we still have ${\delta'}^2\gamma \neq 0$. Nevertheless, we can also remedy this by introducing an auxiliary field $H'$ such that 
\begin{equation}
    \delta' \gamma = H', \qquad \delta' H'=0,
\end{equation}
where the equation of motion for $H'$ will mean that it is equal to the variation of $\eta$.

Letting 
\begin{equation}\label{top-hol gauge fermion}
    \begin{aligned}
    V'_1 &= \lambda_{4w\bar{z}}\big(F_{\bar{w}z}-tD_{\bar{w}}B_{z} \big)\,,\\
    V'_2 &= it\bar{\lambda}_{3wz}\big(F_{\bar{w}\bar{z}}+t^{-1}D_{\bar{w}}B_{\bar{z}} \big)\,,\\
    V'_3 &= g^{z\bar{z}}\eta\big(F_{z\bar{z}}-i[B_{z}, B_{\bar{z}}]+tD_{\bar{z}}B_{z}+t^{-1}D_{z}B_{\bar{z}}\big) \,,\\
    V'_4 &= -\gamma\bigg(\frac{1}{2}H'-t^{-1}D_{z}B_{\bar{z}}+tD_{\bar{z}}B_{z}- g^{w\bar{w}}\big(F_{w\bar{w}}+i[B_{\bar{w}}, B_{w}]\big)+2i[C, C^{\dagger}]\bigg)\,,\\
    V'_5 &= g^{w\bar{w}}D_{\bar{w}}C^{\dagger}\big(\bar{\lambda}_{2w}+it\lambda_{1w} \big)+itg^{z\bar{z}}\big(D_{\bar{z}}C^{\dagger}-it^{-1}[B_{\bar{z}}, C^{\dagger}]\big)\lambda_{1z}\\
    &+g^{z\bar{z}}\big(D_{z}C^{\dagger} +it[C^{\dagger}, B_{z}] \big)\bar{\lambda}_{2\bar{z}}\,,\\
    V'_6 &= i(\bar{\lambda}_1-t\lambda_2)D_{\bar{w}}B_{w} +tg^{z\bar{z}}g^{w\bar{w}}\lambda_{2\bar{w}z}\big(D_{\bar{z}}B_{w}+i [B_{w}, B_{3\bar{z}}] \big)\\
    &+ ig^{z\bar{z}}g^{w\bar{w}}\bar{\lambda}_{1\bar{w}\bar{z}}\big(D_{z}B_{w} -t[B_{z},B_{w}]\big)\,,\\
    V'_7 &= -g^{w\bar{w}}[B_{w}, C^{\dagger}](\lambda_{3\bar{w}}+it\bar{\lambda}_{4\bar{w}}),
    \end{aligned}
\end{equation}

the action in \eqref{4d susy action} can be written in the form
\begin{equation}\label{top-hol q exact}
    S = \frac{1}{e^2}\sum_i^{7}\int_{M_4}d^2z d^2w \sqrt{g}\,\text{Tr}\, \{{\cal Q}', V'_i \} -\frac{i\tau}{4\pi}\int_{M_4} \text{Tr}F \wedge F+\dots
\end{equation}
where $``\dots$" represents terms which are \emph{not} $\mathcal{Q}'$-exact. 

Note that the terms in $``\dots$" cannot contain the metric on $\Sigma_1$. This is because the theory is topological along $\Sigma_{1}$, so any variation of the action with respect to its metric must at most be $\mathcal{Q}'$-exact, but these terms are not. In other words, the path integral of the theory will be independent of the metric on $\Sigma_1$ even in the presence of the terms in ``$\dots$'', as expected. 

%$g^{z\bar{z}}$, $g^{zz}$ or  $g^{\bar{z}\bar{z}}$ \footnote{In complex coordinates, $g^{zz}=g^{\bar{z}\bar{z}}=g^{ww}=g^{\bar{w}\bar{w}}=0$. So, it won't contain these metric components either.}

%Localization of the path integral onto a moduli space of field configurations, like in the case of the previous theory that is topological on all directions of $M_4$, depend on the fact that the theory is independent of the coupling constant, whereby weak-coupling approximations are in fact, exact. Since the action is of the form \eqref{top-hol q exact}, where there are terms that are not $\mathcal{Q}'$-exact, it is not immediately clear that the weak-coupling approximation can be taken, since the coupling $e$ is present in ``$\dots$'' as well. However, we note that terms in the action containing only bosons can be generated by \eqref{top-hol gauge fermion}. Thus, we can rescale the fermions to remove the dependence on $e$ in ``$\dots$'', whence $e$ appears only in the $\mathcal{Q}'$-exact part of \eqref{top-hol q exact}. We can then apply the argument for localization of the path integral onto a moduli space $\mathcal{M}'$ of classical field configurations.

\bigskip\noindent\textit{A Complexified Gauge Connection on $M_4$}
\vspace*{0.5em}\\ 
In contrast to the topological theory from $\mathcal{Q}$-cohomology earlier, we can introduce a complex gauge connection $\mathcal{A}\in\Omega^1(M_4)$,
\begin{equation} \label{curly A}
    \begin{aligned}
    \mathcal{A}&=\mathcal{A}_{z}dz + \mathcal{A}_{\bar{z}}d\bar{z}+\mathcal{A}_{w}dw+\mathcal{A}_{w}d\bar{w}\\
    &=(A_{z}+tB_{z})dz+(A_{\bar{z}}-t^{-1} B_{\bar{z}})d\bar{z}+A_{w}dw+A_{\bar{w}}d\bar{w}.\\
    \end{aligned}
\end{equation}
%where $B_z = B_{z}$ and $B_{\bar{z}}= B_{\bar{z}}$ for ease of notation.
The transformations in \eqref{4d susy transformations top-hol} involving the complexified field strength are (after suitable rescalings)
\begin{equation}
    \begin{aligned}
    \delta'\eta &= \epsilon_{4}\mathcal{F}_{z\bar{z}}g^{z\bar{z}},\\
    \delta'\bar{\lambda}_{3wz} &=-it^{-1}\epsilon_{4} \mathcal{F}_{wz},\\
    \delta' \lambda_{4w\bar{z}}&=\epsilon_{4}\mathcal{F}_{w\bar{z}},
    \end{aligned}
\end{equation}
where $\mathcal{F}\in\Omega^2(M_4)$ is the complexified two-form gauge field strength on $M_4$.

\bigskip\noindent\textit{The BPS Equations}
\vspace*{0.5em}\\
Consequently, from $\delta \psi=0$ (where $\psi$ represents a generic fermionic field) which minimises the action in \ref{top-hol q exact}, the corresponding BPS equations are:
\begin{subequations}\label{4d BPS top-hol}
   \begin{empheq}[box=\widefbox]{align}
     %\mathcal{V}_{\beta}&=0\label{4d BPS a}\\
   %\mathcal{F}_{\bar{z}z}&=0\label{4d BPS top-hol b}\\
      \mathcal{F}_{\bar{z}z}=\mathcal{F}_{wz}=\mathcal{F}_{w\bar{z}}&=0 \label{4d BPS top-hol a}\\
   D_{\bar{z}}B_{\bar{w}}-it^{-1}[B_{\bar{z}}, B_{\bar{w}}]&=0 \label{4d BPS top-hol b}\\
    D_{z}B_{\bar{w}}+it[B_{z}, B_{\bar{w}}]&=0 \label{4d BPS top-hol c}\\
   D_{\bar{z}}C-it^{-1}[ B_{\bar{z}}, C]&=0 \label{4d BPS top-hol d}\\
    D_{z}C+it[B_{z},C]&=0 \label{4d BPS top-hol e}\\
    D_{w}B_{\bar{w}}=D_{w}C&=0 \label{4d BPS top-hol f}\\
    %g^{w\bar{w}}\big(F_{w\bar{w}}+i[B_{\bar{w}}, B_{w}]\big)-i[C, C^{\dagger}]&=0\label{4d BPS top-hol i}\\
     [B_{\bar{w}}, C]=H'&=0\label{4d BPS top-hol g}
    %itD_{\bar{z}}B_{z}-it^{-1}D_{z}B_{\bar{z}}-2i(F_{w\bar{w}}+i[B_{\bar{w}}, B_{w}])+2[C, C^{\dagger}]&=0 
\end{empheq} 
\end{subequations}
Configurations of bosonic fields satisfying \eqref{4d BPS top-hol} constitute a \emph{novel} moduli space $\mathcal{M}'$ that the path integral localizes on.\footnote{The terms in ``$\dots$'' of the action \ref{top-hol q exact} do contain the inverse metric on $\Sigma_2$. However, as the theory is topological along $\Sigma_1$, one can scale up $\Sigma_1$, where this can alternatively be understood as scaling down $\Sigma_2$. In turn, this means that the terms in ``$\dots$'' will scale up, whence contributions to the path integral would continue to be dominated by bosonic field configurations which obey \ref{4d BPS top-hol} that minimize the action via minimizing its ${\cal Q}'$-exact part.}

\bigskip\noindent\textit{$\tau$-independent Correlation Functions}
\vspace*{0.5em}\\
One can show that unlike the fully-topological theory, correlation functions in this topological-holomorphic theory do not depend on $\tau$. Let us explain this now. 

Firstly, the fact that we are able to express the supersymmetry variations in terms of a complexified connection $\mathcal{A}$ means that we are able to add $\mathcal{Q}'$-exact terms to the topological term such that we have (see for example, \cite{ashwinkumar2019boundary}) 
\begin{equation}\label{curly instanton}
    \frac{i\tau}{4\pi}\int_{M_4} \text{Tr}F \wedge F \sim \frac{i\tau}{4\pi}\int_{M_4} \text{Tr}\mathcal{F} \wedge \mathcal{F},
\end{equation}
where $\sim$ in the above indicates that both the left- and right-hand side are $\mathcal{Q}'$-cohomologous, i.e., they are equivalent in the path integral. 

%This implies that the connections $A$ and $\mathcal{A}$ have the same instanton number.

Secondly, this is a topological term, where only the zero-modes of the fields are present. From the BPS equation \eqref{4d BPS top-hol a} that the zero modes obey, which states that  $\mathcal{F}_{wz}=\mathcal{F}_{w\bar{z}}=\mathcal{F}_{z\bar{z}}=0$, it would mean that $\mathcal{F}\wedge \mathcal{F}=0$ . Hence, the theory will not have any dependence on $\tau$, and neither will there be non-trivial instanton configurations.

\bigskip\noindent\textit{Purely Perturbative Holomorphic Correlation Functions}
\vspace*{0.5em}\\
Correlation functions will thus be given by
\begin{equation}\label{4d top-hol corr function}
    \langle \Pi_i\mathcal{O}'_i \rangle_{4d}\big(w, G\big) = \int_{\mathcal{M}'}D\phi'\; \Pi_i\mathcal{O}'_i e^{-S},
\end{equation}
where $D\phi'$ represents the integration measure over all field configurations, and $\mathcal{O}'_i=\mathcal{O}'_i(w)$ is a local operator in $\mathcal{Q}'$-cohomology that must therefore depend on $w$, i.e., the correlation functions are holomorphic in $w$.  The ghost numbers of the fields are the same as before, and non-vanishing correlation functions have to satisfy an anomaly cancellation condition as well. 

%The correlators of $\mathcal{Q}'$-cohomology operators however, have a dependence on the $w$-coordinate, as can be seen from \eqref{susy algebra top-hol}, where correlation functions are holomorphic in $w$. 

Since there will not be any non-perturbative instanton contributions to the path integral, the holomorphic  correlations functions can be understood to be purely perturbative. This is an important observation that will allow us to make contact with the mathematical theory of CDOs in the next section.    

\section{An $\mathcal{N}=(4,4)$ Holomorphic Sigma-Model on $\Sigma_2$ and a 4d-2d Correspondence}\label{section: holomoprhic sigma model 2d}  

In this section, we again perform a dimensional reduction of the 4d theory along $\Sigma_{1}$ to obtain an effective 2d theory on $\Sigma_2$. Note that we can do this because the 4d theory is still topological along $\Sigma_1$. 

\bigskip\noindent\textit{Dimensional Reduction Along $\Sigma_1$}
\vspace*{0.5em}\\
A similar analysis that led us to \ref{finiteness condition topological} tells us that from \eqref{4d BPS top-hol a}, we have the finiteness condition
\begin{equation}
    \mathcal{F}_{z\bar{z}}=F_{z\bar{z}}-i[B_{z},B_{\bar{z}}]+tD_{z}B_{\bar{z}}+t^{-1}D_{\bar{z}}B_{z}=0.
\end{equation}
This can be solved by 
\begin{subequations}\label{hitchin equation top hol}
\begin{empheq}[box=\widefbox]{align}
  F_{z\bar{z}}-i[B_{z},B_{\bar{z}}]&=0\\
    D_{\bar{z}}B_{z}&=0
\end{empheq}
\end{subequations}
which again, are Hitchin's equations. Like earlier, the fields $(A_{\Sigma_{1}}, B_{\Sigma_{1}})$, which are solutions to the above Hitchin's equations, define a map $\Phi: \Sigma_2 \to \mathcal{M}^G_H(\Sigma_{1})$. Similar to before, we set $\Sigma_{1}$ to have genus $g\geq 2$ for simplicity. 

\bigskip\noindent\textit{The Surviving 4d Action}
\vspace*{0.5em}\\
Also, the surviving 4d action (excluding the topological term) is the same as that given in \eqref{eff 2d action}, since the dimensional reduction along $\Sigma_1$ is independent of the choice of the cohomology of scalar supercharges. We restate it here for later convenience:
\begin{equation}\label{S-surv}
    \begin{aligned}
    S_{\text{surv}} &= \frac{1}{e^2}\int_{M_4} d^2z d^2w \sqrt{g}\,\text{Tr}\bigg[\frac{1}{2}(F_{zw}F^{zw}+F_{\bar{z}w}F^{\bar{z}w}+F_{z\bar{w}}F^{z\bar{w}}+F_{\bar{z}\bar{w}}F^{\bar{z}\bar{w}})+g^{z\bar{z}}g^{w\bar{w}}(D_{w}B_{z}D_{\bar{w}}B_{\bar{z}}\\
    &+D_{\bar{w}}B_{z}D_{w}B_{\bar{z}}
    -i\bar{\lambda}_{1\bar{w}\bar{z}}D_{w}\lambda_{1z}-i\bar{\lambda}_{2\bar{z}}D_{w}\lambda_{2\bar{w}z}-i\bar{\lambda}_{3wz}D_{\bar{w}}\lambda_{3\bar{z}}-i\bar{\lambda}_{4z}D_{\bar{w}}\lambda_{4w\bar{z}})+\dots\bigg]
    \end{aligned}
\end{equation}

\bigskip\noindent\textit{An ${\cal N} = (4,4)$ Holomorphic Sigma Model on $\Sigma_2$}
\vspace*{0.5em}\\
That the fields $(A_{\Sigma_{1}}, B_{\Sigma_{1}})$ define a map $\Phi: \Sigma_2 \to \mathcal{M}^G_H(\Sigma_{1})$ means that the 4d theory should reduce to a 2d sigma model on $\Sigma_2$ with  hyperK\"{a}hler target $\mathcal{M}^G_H(\Sigma_{1})$. As the target is hyperK\"{a}hler, the sigma model will have underlying ${\cal N} = (4,4)$ supersymmetry, as before. Moreover, from \eqref{top charge hol in w}, we know that the (${\cal Q}'$-cohomology of the) sigma-model must be holomorphic in $w$. Altogether, this means that we have an ${\cal N} = (4,4)$ holomorphic sigma model on $\Sigma_2$ with target $\mathcal{M}^G_H(\Sigma_{1})$.

\bigskip\noindent\textit{The Complex Structure of the Sigma Model}
\vspace*{0.5em}\\
For a general $t\neq 0, \infty$, \eqref{hitchin equation top hol} implies flat complexified connections on $\Sigma_{1}$, whence $\mathcal{M}^G_H(\Sigma_{1})$ can be identified as $\mathcal{M}^{G_{\mathbb{C}}}_{\text{flat}}(\Sigma_{1})$, the moduli space of flat complexified connections on $\Sigma_{1}$, where the complex structure of the sigma model must therefore be a linear combination of complex structures $J$ and $K$.

\bigskip\noindent\textit{The 2d BPS Equations}
\vspace*{0.5em}\\
Let us now determine the 2d BPS equations of the sigma model. Before we proceed further, note that \eqref{4d BPS top-hol b} and \eqref{4d BPS top-hol d} indicate that $B_{\bar{w}}$ and $C$, as scalars on $\Sigma_{1}$, are covariantly constant with respect to the complexified gauge connection $\mathcal{A}_{\Sigma_{1}}$. Hitchin's equations imply that $\mathcal{A}_{\Sigma_{1}}$ is flat whence we can go to pure gauge and set it to zero, and since we want irreducible gauge connections, we can only have constant (zero mode) solutions $B_{\bar{w}}=C=0$.

Now, let us dimensionally reduce the 4d BPS equations \ref{4d BPS top-hol} to get the 2d BPS equations. We need only consider \eqref{4d BPS top-hol a}, since it is the only one that survives after setting  $B_{\bar{w}}=C=0$, which moreover relates to the terms which appear in \eqref{S-surv} (since it involves mixed coordinates), whence it would lead to the relevant 2d BPS equations. As before, we can take $F_{wz}=\partial_{w}A_{z}$ and $D_{w}B_{z}=\partial_{\bar{w}}B_z$ etc., since $A_{{w}}$ does not have derivatives on $\Sigma_2$ and is thus a non-dynamical field that can be integrated out via its equation of motion (which expresses it as a combination of fermionic fields).

From \eqref{4d BPS top-hol a}, we then obtain the 2d BPS equations of the sigma model as
\begin{subequations}\label{2d bps top-hol equation}
\begin{empheq}[box=\widefbox]{align}
     \partial_{w}(A_{z}+t B_{z})&=0\label{2d bps equation top-hol a}\\
     \partial_{w}(A_{\bar{z}}-t^{-1}B_{\bar{z}})&=0\label{2d bps equation top-hol b}
\end{empheq}
\end{subequations}
The first equation tells us that the map $\Phi: \Sigma_2 \to \mathcal{M}^G_H(\Sigma_{1})$ is anti-holomorphic if we choose a complex structure on $\mathcal{M}^G_H(\Sigma_{1})$ in which $A_{z}+tB_{z}$ is a holomorphic coordinate. Similarly, the second equation says that the map $\Phi$ is anti-holomorphic if a complex structure on $\mathcal{M}^G_H(\Sigma_{1})$ is chosen such that $A_{\bar{z}}-t^{-1}B_{\bar{z}}$ is a holomorphic coordinate. These are exactly $\mathcal{A}_{z}$ and $\mathcal{A}_{\bar{z}}$ defined in \eqref{curly A}, respectively.

\bigskip\noindent\textit{The Moduli Space that the 2d Path Integral Localizes On}
\vspace*{0.5em}\\
Let us now rewrite some of \eqref{4d susy transformations top-hol} using $\mathcal{A}_{z}$ and $\mathcal{A}_{\bar{z}}$ (where we henceforth set $\epsilon_4=1$ for convenience):
\begin{equation}\label{curly A transform}
    \begin{aligned}
    \delta' \mathcal{A}_{z}&=0, \quad & \delta' \mathcal{A}_{\bar{z}}&=0,\\
    \delta' \bar{\mathcal{A}}_{\bar{z}}&=(1+|t|^2)\lambda_{3\bar{z}}, \quad & \delta' \bar{\mathcal{A}}_{z}&=i(t+\bar{t}^{-1})\bar{\lambda}_{4z}.
    \end{aligned}
\end{equation}
Notice that we have the following correspondences between the 4d fields and the 2d sigma model fields
\begin{equation}\label{field corres top-hol}
    \begin{aligned}
    \mathcal{A}_{\bar{z}} 	&\leftrightarrow X^i, &\qquad \mathcal{A}_{z} 	&\leftrightarrow Y^i,\\
    \bar{\mathcal{A}}_{z}&\leftrightarrow X^{\bar{i}}, &\qquad  \bar{\mathcal{A}}_{\bar{z}}&\leftrightarrow Y^{\bar{i}},\\
    \lambda_{3\bar{z}}&\leftrightarrow \psi_{1}^{\bar{i}}, &\qquad \bar{\lambda}_{4z}&\leftrightarrow\psi^{\bar{i}}_{2},\\
    \bar{\lambda}_{3wz}&\leftrightarrow\rho_{1w}^{i}, &\qquad \lambda_{4w\bar{z}}&\leftrightarrow\rho^{i}_{2w},
    \end{aligned}
\end{equation}
which, from \eqref{curly A transform}, will allow us to write the transformations for the bosonic fields in the sigma model as
\begin{equation}\label{2d susy 4,4 boson 1 top-hol}
    \begin{aligned}
    \delta' X^{i} &= 0, \quad & \delta' Y^{i} &=0,\\
    \delta' X^{\bar{i}}&= \psi_{1}^{\bar{i}}, \quad & \delta' Y^{\bar{i}}&= \psi_{2}^{\bar{i}}
    \end{aligned}
\end{equation}
(after appropriately rescaling the fermionic fields), where $i, \bar{i}=1,2,\dots, \frac{1}{2}\text{dim}_{\mathbb{C}}\big(\mathcal{M}^G_H(\Sigma_{1}) \big)$. 

From \eqref{4d susy transformations top-hol}, the variations of the above fermionic fields in the sigma model are
\begin{equation}\label{2d susy 4,4 fermion 1 top-hol}
    \begin{aligned}
    \delta' \psi_{1}^{\bar{i}} &=0, \quad &  \delta' \psi_{2}^{\bar{i}}&=0,\\
    \delta' \rho^{i}_{1w} &=  \partial_{w}X^{i}+\dots, \quad& \delta' \rho_{2w}^{i}&= \partial_{w}Y^{i}+\dots,
    \end{aligned}
\end{equation}
where ``$\dots$" representa a term involving the curvature of $\mathcal{M}^G_H(\Sigma_{1})$ coupled to fermions, which is standard in 2d supersymmetric sigma models.

For completeness, let also state the correspondences between the rest of the 4d fermionic fields and the rest of the 2d sigma model fermionic fields
\begin{equation}\label{field corres 2 top-hol}
    \begin{aligned}
    \lambda_{1z}&\leftrightarrow \psi_{1}^{i}, &\qquad \bar{\lambda}_{2\bar{z}}&\leftrightarrow\psi^{i}_{2},\\
    \bar{\lambda}_{1\bar{w}\bar{z}}&\leftrightarrow\rho_{1\bar{w}}^{\bar{i}}, &\qquad \lambda_{2\bar{w}z}&\leftrightarrow\rho^{\bar{i}}_{2\bar{w}},
    \end{aligned}
\end{equation}
whereby they vary like in \eqref{2d susy 4,4 fermion 1 top-hol}. 

From \eqref{2d susy 4,4 boson 1 top-hol}, $\mathcal{Q}'$ appears to be a Dolbeault operator on $\mathcal{M}^G_H(\Sigma_{1})$. This interpretation however, will not always hold in a holomorphic sigma-model as shown in \cite{tan2006two, witten2007two}. %This fact will be discussed further when we examine the correlation functions of the sigma-model.

 Finally, recalling that in the present case, $\mathcal{M}^G_H(\Sigma_{1}) = \mathcal{M}^{G_{\mathbb{C}}}_{\text{flat}}(\Sigma_{1})$,  let us write the fields in the sigma-model as
\begin{equation}\label{M_flat combine vector}
    \begin{aligned}
    Z^I &= X^i\oplus Y^i, &\qquad Z^{\bar{I}} &= X^i\oplus Y^i,\\
    \psi^{I} &= \psi^{i}_{1} \oplus \psi^{i}_{2}, &\qquad \psi^{\bar{I}} &= \psi^{\bar{i}}_{1} \oplus \psi^{\bar{I}}_{2},\\
     \rho^{I}_{w} &= \rho^{i}_{1w} \oplus \rho^{i}_{2w}, &\qquad \rho^{\bar{I}}_{\bar{w}} &= \rho^{\bar{i}}_{1\bar{w}} \oplus \rho^{\bar{i}}_{2\bar{w}},
    \end{aligned}
\end{equation}
where $I, \bar{I}=1,2,\dots, \text{dim}_{\mathbb{C}}\big(\mathcal{M}^{G_{\mathbb{C}}}_{\text{flat}}(\Sigma_{1}) \big)$. In contrast to the previous case of the topological $A$-model where we are strictly in complex structure $I$ (for $s\neq 0, \infty$) and $\mathcal{M}^G_{\text{Higgs}}(\Sigma_{1}) $ is distinctly split into a base and fiber structure with equal dimensions, the same cannot be said of $\mathcal{M}^{G_{\mathbb{C}}}_{\text{flat}}(\Sigma_{1})$. It is thus more convenient to express the fields as in \eqref{M_flat combine vector}.

From \eqref{2d susy 4,4 fermion 1 top-hol}, the BPS equations can also be written as\footnote{The 2d BPS equations are defined by also setting the variation of bosonic fields to zero. This results in setting (the fermions in) the ``$\dots$'' terms of \eqref{2d susy 4,4 fermion 1 top-hol} to zero.}
\begin{equation}
    %\boxed{\partial_{w}X^{i}=\partial_{w}Y^{i}=0}
    \boxed{\partial_{w}Z^{I}=0}
\end{equation}
and the map $Z: \Sigma_2 \to \mathcal{M}^{G_{\mathbb{C}}}_{\text{flat}}(\Sigma_{1})$ is anti-holomorphic.
The path integral thus localizes to an integral over a moduli space of anti-holomorphic maps
\begin{equation} \label{Mmaps top-hol}
    \boxed{\mathcal{M}_{\text{maps}}' = \{ Z:\Sigma_2\to\mathcal{M}^{G_{\mathbb{C}}}_{\text{flat}}(\Sigma_{1}) \;|\; \partial_{w}Z^{I}=0 \}}
\end{equation}

\subsection{The \u{C}ech Cohomology of the Sheaf of CDOs and a Quasi-topological String}
\bigskip\noindent\textit{$\tau$-independent Correlation Functions}
\vspace*{0.5em}\\
Let us first address the dependency of correlation functions on $\tau$.
Since it was established that the complex structure is $J+ \alpha K$ for $t\neq 0, \infty$, the topological worldsheet instanton term will be of the form
\begin{equation}
    i\tau\int_{C}\Phi^{*}(\omega_t),
\end{equation}
where $\omega_t$ is the K\"{a}hler form of Hitchin's moduli space in complex structure $J+\alpha K$, where $\alpha \in \mathbb{C}$. Since $\omega_J$ and $\omega_K$ and hence, $\omega_t$, are $\mathcal{Q}'$-exact,
%they are exact with respect to the variation of $(A_{\Sigma_{1}}, B_{\Sigma_{1}})$ on $\mathcal{M}^G_H(\Sigma_{1})$. But the variation along the target space is generated by $\mathcal{Q}'$. 
the topological worldsheet instanton term above will not contribute to the path integral, whence the 2d correlation functions will not depend on $\tau$. %\footnote{It was shown in Section 4.1 of \cite{kapustin2006electric} that $\omega_J$ and $\omega_K$ are exact wrt the exterior derivative on $\mathcal{W}$, the infinite-dimensional affine space spanned by $(A_{\Sigma_1}, B_{\Sigma_1})$.} 

\bigskip\noindent\textit{A Purely-Perturbative Sigma Model}
\vspace*{0.5em}\\
This is consistent with the analysis of the topological term in the 4d theory, where it was also shown that 4d correlation functions have no dependence on $\tau$. Furthermore, just like in the 4d case, the 2d correlation functions are purely-perturbative, i.e., they do not receive any instanton corrections. Last but not least, as there are no contributions from worldsheet instantons, the path integral effectively localizes to degree-zero maps in \eqref{Mmaps top-hol}, i.e, the effective target space of the purely-perturbative sigma model is always $\mathcal{M}^{G_{\mathbb{C}}}_{\text{flat}}(\Sigma_{1})$.

\bigskip\noindent\textit{The ${\cal Q}'$-cohomology and the Sheaf of CDOs}
\vspace*{0.5em}\\
The transformations of the fields in \eqref{2d susy 4,4 boson 1 top-hol}, \eqref{2d susy 4,4 fermion 1 top-hol} are like those of the half-twisted $A$-model studied in \cite{tan2006two, witten2007two}, albeit for an anti-holomorphic version and a doubling of the number of fermion fields due to a higher $\mathcal{N}=(4,4)$ supersymmetry. 

The dependence on $w$ only means that observables should be understood as local \emph{holomorphic} operators. Moreover, correlation functions of such operators are purely perturbative, as explained above. These features of our sigma model also coincide with the features of the half-twisted $A$-model studied in \cite{tan2006two, witten2007two}.

%The mathematical description of these observables, however, are more complex as compared to that in a topological theory. This is because 

As mentioned earlier, the interpretation of $\mathcal{Q}'$ as a Dolbeault operator might not always hold in a holomorphic theory. Thus, while the $\mathcal{Q}$-cohomology of our earlier 2d topological $A$-model can simply be described by ordinary de Rham cohomology, the $\mathcal{Q}'$-cohomology of our present 2d holomorphic sigma model cannot be described by ordinary Dolbeault cohomology. Nevertheless, from the preceding two paragraphs, it would mean that the analysis in~\cite{tan2006two, witten2007two} applies to our holomorphic sigma model, whence we can conclude  
that the \u{C}ech cohomology of the sheaf $\Omega_{\text{cdo}}$ of chiral differential operators (CDOs) on $\mathcal{M}^{G_{\mathbb{C}}}_{\text{flat}}(\Sigma_{1})$ is the proper framework to describe the $\mathcal{Q}'$-cohomology. Specifically, local holomorphic observables $\tilde{\mathcal{O}}'$ in the $\cal Q'$-cohomology can be identified with \u{C}ech cohomology classes
\begin{equation}
    \tilde{\mathcal{O}}' \cong H^{*}_{\text{\u{C}ech}}(\mathcal{M}^{G_{\mathbb{C}}}_{\text{flat}}(\Sigma_{1}), \Omega_{\text{cdo}}),
\end{equation}
where, due to the \u{C}ech-Dolbeault isomorphism, the degree of $H^{*}_{\text{\u{C}ech}}$ is given by the number of $\psi^{\bar{I}}$ in $\tilde{\mathcal{O}}'$.

%In general, it is possible to have CDOs on the moduli space that the sigma-model localizes on ie, CDOs on $\mathcal{M}'_{\text{maps}}$. In this case however, since contributions from the topological worldsheet instanton term are trivial, we will only have degree-zero maps of  $Z:\Sigma_2\to \mathcal{M}^{G_{\mathbb{C}}}_{\text{flat}}(\Sigma_{1})$, where $\mathcal{M}'_{\text{maps}}$ is simply $\mathcal{M}^{G_{\mathbb{C}}}_{\text{flat}}(\Sigma_{1})$.

\bigskip\noindent\textit{Purely Perturbative Holomorphic Correlation Functions}
\vspace*{0.5em}\\
Thus, the purely-perturbative holomorphic correlation functions will be given by
\begin{equation}\label{2d top-hol corr function}
    \langle \Pi_i\tilde{\mathcal{O}}'_i \rangle_{2d}\big(w, \mathcal{M}^{G_{\mathbb{C}}}_{\text{flat}}(\Sigma_{1})\big) = \int_{\mathcal{M}^{G_{\mathbb{C}}}_{\text{flat}}}D\tilde{\phi}'\; \Pi_i\tilde{\mathcal{O}}'_i e^{-S} = \int_{\mathcal{M}_{\text{flat}}^{G_{\mathbb{C}}}(\Sigma_1)}  \bigotimes_i H^*_{\text{\u{C}ech}, i},
\end{equation}
where $D\tilde{\phi}'$ represents the integration measure over all field configurations, and $\tilde{\mathcal{O}}'_i=\tilde{\mathcal{O}}'_i(w)$ is an observable in the $\mathcal{Q}'$-cohomology. Note that a ghost number anomaly condition has to be satisfied, as in the 4d theory, and this will result in the product of a ``top class'' in the third expression above. 

\bigskip\noindent\textit{A Quasi-topological String}
\vspace*{0.5em}\\
Note that the holomorphic sigma model is also a quasi-topological sigma model - from the anaylsis in~\cite{tan2006two,witten2007two} which applies here, one can compute that only the antiholomorphic stress tensor is $\cal Q'$-exact. As such, it defines not a topological but a quasi-topological string. In turn, this means that \eqref{2d top-hol corr function} can be interpreted as a contribution to the scattering amplitude of the quasi-topological string at genus $g$ of $\Sigma_2$, where the string vertex operators will be given by the $\tilde{\mathcal{O}}'_i$'s which can be identified with the classes $ H^{*}_{\text{\u{C}ech}}(\mathcal{M}^{G_{\mathbb{C}}}_{\text{flat}}(\Sigma_{1}), \Omega_{\text{cdo}})$.

\subsection{A 4d-2d Correspondence between Holomorphic Invariants of $M_4$ 
and CDOs on the Space of Flat Complexified Connections}

As before, we have a 4d-2d correspondence from the topological-holomorphic twist which renders the theory topological along $\Sigma_{1}$:
\begin{equation}\label{4d-2d top-hol corr}
    \boxed{ \langle \Pi_i\mathcal{O}'_i \rangle_{4d}\big(w, G\big)= \int_{\mathcal{M}_{\text{flat}}^{G_{\mathbb{C}}}(\Sigma_1)} \bigotimes_i H^*_{\text{\u{C}ech}, i}  =  \text{CDO}\big(w, \mathcal{M}^{G_{\mathbb{C}}}_{\text{flat}}(\Sigma_{1})\big)}
\end{equation}
where $\text{CDO}\big(w, \mathcal{M}^{G_{\mathbb{C}}}_{\text{flat}}(\Sigma_{1})\big)$ is an evaluation over $\mathcal{M}^{G_{\mathbb{C}}}_{\text{flat}}(\Sigma_{1})$ of a  product of the classes $H^{*}_{\text{\u{C}ech}}$ that correspond to the operators $\tilde{\mathcal{O}}'_{i}(w)$.

We thus have a correspondence between holomorphic invariants of $M_4$ given by the LHS of \eqref{4d-2d top-hol corr}, and CDOs on the space of flat complexified connections given by the RHS of \eqref{4d-2d top-hol corr}. 

\section{A Holomorphic Theory on $M_4$}\label{section: holomorphic m4}

In this section, we briefly consider the case of $t=0$ in $\mathcal{Q}'$-cohomology. We now have just a single scalar supercharge $\bar{\mathcal{Q}}_3$, where the (twisted) supersymmetry algebra \eqref{susy algebra top-hol} tells us that
\begin{equation}\label{susy algebra t=0}
    \{\bar{\mathcal{Q}}_3, \mathcal{Q}_{3\bar{w}} \} \propto P_{\bar{w}}, \qquad \{\bar{\mathcal{Q}}_{3}, \mathcal{Q}_{3\bar{z}} \} \propto P_{\bar{z}}.
\end{equation}
The theory is thus fully-holomorphic on $M_4=\Sigma_{1} \times \Sigma_2$, where correlation functions can have a dependence on $z$ and $w$.

\bigskip\noindent\textit{The BPS Equations}
\vspace*{0.5em}\\
Let us ascertain the BPS equations in this case where $t=0$. 
%In particular, we wish to examine how taking $t=0$ affects the dependence of correlation functions on $\tau$. 
%We thus only focus on BPS equations involving $A_{\mu}$ and $B_z, B_{\bar{z}}$, since these are the fields that constitute $\mathcal{A}$. 
Letting $t=0$ is equivalent to setting $\epsilon_4=0$ in \eqref{4d susy transformations top-hol}. The BPS equations are then
%\begin{equation}\label{4d bps single scalar charge 1}
%    \begin{aligned}
%    D_{w}B_{\bar{z}}&=0\\
%    F_{wz}&=0\\
%    D_{\bar{z}}B_{z}&=0
%    \end{aligned}
%\end{equation}
\begin{subequations}\label{4d bps single scalar charge 1}
   \begin{empheq}[box=\widefbox]{align}
   g^{z\bar{z}}\big(F_{z\bar{z}}-i[B_{z}, B_{\bar{z}}]\big)+
    g^{w\bar{w}}\big(F_{w\bar{w}}+i[B_{\bar{w}}, B_{w}]\big)-i[C, C^{\dagger}]&=0\label{4d BPS single hol a}\\  
  D_{w}C=D_{z}C= g^{w\bar{w}}D_{w}B_{\bar{w}}=D_{z}B_{\bar{w}}=D_{w}B_{\bar{z}}=g^{z\bar{z}}D_{z}B_{\bar{z}}&=0\label{4d BPS single hol b}\\  
    F_{wz}=[B_{\bar{z}}, B_{\bar{w}}]=[B_{\bar{z}}, C]= [B_{\bar{w}}, C]&=0 \label{4d BPS single hol c} 
\end{empheq} 
\end{subequations}
%and 
%\begin{equation}\label{4d bps single scalar charge 2}
 %%   g^{z\bar{z}}(F_{z\bar{z}}-i[B_{z},B_{\bar{z}}])
 %   -g^{w\bar{w}}\big(F_{w\bar{w}}+i[B_{\bar{w}}, B_{w}]\big)=0
%\end{equation}
%Together with the remaining BPS equations from \eqref{4d BPS top-hol} when $t=0$, t
These constitute a \emph{novel} moduli space $\mathcal{M}'_0$ that the path integral localizes on (where the subscript $0$ is to indicate that we are considering $t=0$ in $\mathcal{Q}'$-cohomology).\footnote{The action continues to be of the form in \ref{top-hol q exact}, but the terms in ``$\dots$'' now contain the inverse metric on $\Sigma_1$ and $\Sigma_2$. However, as the theory is still scale-invariant along the Riemann surfaces $\Sigma_1$ and $\Sigma_2$, one can scale them down whence the terms in ``$\dots$'' will scale up, and contributions to the path integral would continue to be dominated by bosonic field configurations which obey \ref{4d bps single scalar charge 1} that minimize the action via minimizing its ${\cal Q}'$-exact part.} 

\bigskip\noindent\textit{Fully-Holomorphic Correlation Functions}
\vspace*{0.5em}\\
From \eqref{4d bps single scalar charge 1}, we see that not all the components of $F_{\mu\nu}$ vanish. Hence, in contrast to the case where $t\neq0, \infty$, we will not always have $\int \text{Tr} F \wedge F = 0$. There can be non-trivial instanton contributions, and correlation functions will have a $\tau$ dependence. Therefore, we will have fully-holomorphic correlation functions of the form
\begin{equation}\label{4d top-hol corr function single scalar charge}
    \langle \Pi_i\mathcal{O}'_{i} \rangle_{4d,0}\big(w,z, \tau, G\big) = \int_{\mathcal{M}_{0}'}D\phi'\; \Pi_i\mathcal{O}'_{0,i} e^{-S}.
\end{equation}

\section{Langlands Duality of Topological and Holomorphic Invariants,
and Mirror Symmetry}\label{section: langlands duality}

With $\mathcal{N}=4$ supersymmetry, the 4d theory possesses an $SL(2, \mathbb{Z})$ symmetry, having both $T$- and $S$-duality. These symmetries act on the complex parameter $\tau$. $T$-duality is simply a symmetry of the theory under a shift $\tau \to \tau+1$. $S$-duality however, relates a theory with gauge group $G$ and coupling $\tau$ to a dual theory with Langlands dual gauge group ${^LG}$ and dual coupling 
\begin{equation}
    {^L\tau} = -\frac{1}{\tau}.
\end{equation}
For a non-simply laced gauge group, the dual coupling is instead 
\begin{equation}
    {^L\tau} = -\frac{1}{n_{\mathfrak{g}}\tau},
\end{equation}
where $n_{\mathfrak{g}}$ is the lacing number of the group. 

\subsection{Langlands Duality in $\mathcal{Q}$-cohomology and Mirror Symmetry of Higgs Bundles}
Up to a possible phase factor of modular weights which is just a constant, under $S$-duality, we have   
\begin{equation}\label{langlands dual 4d top corr}
   \boxed{ \langle \Pi_i\mathcal{O}^{(m)}_i \rangle_{4d}\big(\tau, G\big) \longleftrightarrow \langle \Pi_i\mathcal{O}^{(m)}_i \rangle_{4d}\big(-1/n_{\mathfrak{g}}\tau, {^LG}\big)}
\end{equation}
In other words, we have a Langlands duality of 4d topological invariants of $M_4$.

Topological invariance will then imply, from \eqref{4d-2d corr top} and \eqref{langlands dual 4d top corr}, that we also have
\begin{equation}\label{langlands dual 2d higgs}
   \boxed{  \langle \Pi_i\tilde{\mathcal{O}}^{(p)}_i \rangle_{2d} \big(\tau, \mathcal{M}^G_{\text{Higgs}}(\Sigma_{1})\big) \longleftrightarrow \langle \Pi_i\tilde{\mathcal{O}}^{(p)}_i \rangle_{2d} \big(-1/n_{\mathfrak{g}}\tau, \mathcal{M}^{^LG}_{\text{Higgs}}(\Sigma_{1})\big)}
\end{equation}
where $\mathcal{M}^G_{\text{Higgs}}(\Sigma_{1})$ and $\mathcal{M}^{^LG}_{\text{Higgs}}(\Sigma_{1})$ are mirror manifolds. In other words, we have a duality of GW invariants that can be interpreted as a mirror symmetry of Higgs bundles.

\bigskip\noindent\textit{An Open $A$-model on $\Sigma_2$ and Homological Mirror Symmetry}
\vspace*{0.5em}\\
If $\Sigma_2$ is of the form $\mathbb{R}\times I$, where $I$ is an interval,  we get an \emph{open} $A$-model with target $\mathcal{M}^G_{\text{Higgs}}(\Sigma_1)$.  This furnishes us with a (derived) category of $A$-branes in $\mathcal{M}^G_{\text{Higgs}}(\Sigma_1)$. Since we have an $A$-model in complex structure $I$, we can only have branes that are of type $(A,*,*)$ in $\mathcal{M}^G_{\text{Higgs}}(\Sigma_1)$. Because the $A$-model in complex structure $I$ will map to itself under 4d $S$-duality, it will mean that `$S$-dual' branes are also of type $(A,*,*)$ in $\mathcal{M}^{^LG}_{\text{Higgs}}(\Sigma_1)$. Some examples of these $A$-branes are given in \cite{kapustin2008note}.

Notice that topological invariance along $\Sigma_1$ implies  that 4d $S$-duality would mean the 2d duality 
\begin{equation}\label{2d partition dual brane}
    \mathcal{Z}_{A,\mathscr{B}}\big(\tau, \mathcal{M}^{G}_{\text{Higgs}}(\Sigma_1)\big) 	\longleftrightarrow \mathcal{Z}_{A,\mathscr{B}}\big(-1/n_{\mathfrak{g}}\tau, \,\mathcal{M}^{^LG}_{\text{Higgs}}(\Sigma_1)\big), 
\end{equation}
where $\mathcal{Z}_{A,\mathscr{B}}$ is the partition function of the open $A$-model with branes $\mathscr B$.\footnote{Because $\Sigma_2$ is flat, the index of the theory is $k=0$. In other words, only the partition function with no operator insertions is non-vanishing. } In turn, this implies a homological mirror symmetry 
\begin{equation}\label{dual Cat A brane}
    \boxed{\text{Cat}_{\text{$A$-branes}}\big(\tau, \mathcal{M}^{G}_{\text{Higgs}}(\Sigma_1)\big) \longleftrightarrow
    \text{Cat}_{\text{$A$-branes}}\big(-1/n_{\mathfrak{g}}\tau, \,\mathcal{M}^{^LG}_{\text{Higgs}}(\Sigma_1)\big) }
\end{equation}
of the $\tau$-dependent category of $A$-branes.

\subsection{Langlands Duality in $\mathcal{Q}'$-cohomology and Mirror Symmetry of Quasi-\\topological Strings}
In the case of $t\neq 0, \infty$, we have seen that the holomorphic correlation functions have no $\tau$ dependence. Nonetheless, from $S$-duality, we still have a Langlands duality 
\begin{equation}\label{langlands dual 4d top-hol corr}
   \boxed{ \langle \Pi_i\mathcal{O}'_i \rangle_{4d}\big(w, G\big) \longleftrightarrow \langle \Pi_i\mathcal{O}'_i \rangle_{4d}\big(w, {^LG}\big)}
\end{equation}
of 4d holomorphic invariants. 

Topological invariance along $\Sigma_{1}$, together with \eqref{4d-2d top-hol corr} will then imply a 2d duality
\begin{equation}\label{langlands dual 2d top-hol }
\boxed{ \text{CDO}\big(w, \mathcal{M}^{G_{\mathbb{C}}}_{\text{flat}}(\Sigma_{1})\big)\longleftrightarrow\text{CDO}\big(w, \mathcal{M}^{{^LG}_{\mathbb{C}}}_{\text{flat}}(\Sigma_{1})\big)}
%    \boxed{ \langle \Pi_i\tilde{\mathcal{O}}'_i \rangle_{2d}\big(w, \mathcal{M}^{G_{\mathbb{C}}}_{\text{flat}}(\Sigma_{1})\big)\longleftrightarrow\langle \Pi_i\tilde{\mathcal{O}}'_i \rangle_{2d}\big(w, \mathcal{M}^{{^LG}_{\mathbb{C}}}_{\text{flat}}(\Sigma_{1})\big)}
\end{equation}
That is, we have a mirror symmetry between CDOs and thus quasi-topological strings on $\mathcal{M}^{G_{\mathbb{C}}}_{\text{flat}}(\Sigma_{1})$ and its mirror  $\mathcal{M}^{{^LG}_{\mathbb{C}}}_{\text{flat}}(\Sigma_{1})$.\footnote{See \eqref{4d-2d top-hol corr}) for the full expression of ``CDO(...)''.} 

When $t=0$, from $S$-duality, there will be a Langlands duality
\begin{equation}\label{langlands 4d hol single}
\boxed{\langle \Pi_i\mathcal{O}'_{i} \rangle_{4d,0}\big(w,z, \tau, G\big) \longleftrightarrow \langle \Pi_i\mathcal{O}'_{i} \rangle_{4d,0}\big(w,z, -1/n_{\mathfrak{g}}\tau, {^LG}\big)}
\end{equation}
of 4d fully-holomorphic invariants. 
%and 2d correlation functions
%\begin{equation}\label{langlands 2d hol single}
    %\boxed{\text{CDO}\big(w,\tau,  \mathcal{M}^{G}_{\text{Higgs}}(\Sigma_1)\big) \longleftrightarrow \text{CDO}\big(w,-1/n_{\mathfrak{g}}\tau,  \mathcal{M}^{{^LG}}_{\text{Higgs}}(\Sigma_1)\big)}
%\end{equation}

\section{A Novel Web of Mathematical Relations}\label{section: web}
\begin{figure}
\begin{center}
  \begin{tikzcd}[column sep=-50pt,row sep=35pt]
  \boxed{\text{Cat}_{\text{$A$-branes}}\big(\tau, \mathcal{M}^{G}_{\text{Higgs}}(\Sigma_1)\big)}
  \arrow[rr, leftrightarrow, dashed, "\substack{\text{Homological}\\\text{ mirror symmetry }\\\text{of Higgs}\\\text{ bundles}}"]
  \arrow[d, leftrightarrow,  "\substack{\text{$M_4=\Sigma_{1} \times \mathbb{R}\times I$,}\\\text{$\Sigma_1\to 0$}}"]
  &&\boxed{\text{Cat}_{\text{$A$-branes}}\big(-1/n_{\mathfrak{g}}\tau, \,\mathcal{M}^{^LG}_{\text{Higgs}}(\Sigma_1)\big)}
  \arrow[d, leftrightarrow,  "\substack{\text{$M_4=\Sigma_{1} \times \mathbb{R}\times I$,}\\\text{$\Sigma_1\to 0$}}"]\\
  \boxed{\langle \Pi_i\mathcal{O}_i \rangle_{4d}\big(\tau, G\big)}
  \arrow[dr, leftrightarrow, "\substack{\text{$M_4=\Sigma_{1} \times \Sigma_2$,}\\\text{$\Sigma_1\to 0$}}"]
  \arrow[rr, leftrightarrow, dashed, "\text{Langlands dual}"]
  &&\boxed{\langle \Pi_i\mathcal{O}_i \rangle_{4d}\big(-1/n_{\mathfrak{g}}\tau, {^LG}\big)}
  \arrow[drr, leftrightarrow,  "\substack{\text{$M_4=\Sigma_{1} \times \Sigma_2$,}\\\text{$\Sigma_1\to 0$}}"]\\
  &\boxed{ \langle \Pi_i\tilde{\mathcal{O}}_i \rangle_{2d} \big(\tau, \mathcal{M}^G_{\text{Higgs}}(\Sigma_{1})\big)}
  \arrow[rrr, leftrightarrow, dashed, "\substack{\text{Mirror symmetry}\\\text{of Higgs}\\\text{bundles}}"]
  &&&\boxed{\langle \Pi_i\tilde{\mathcal{O}}_i \rangle_{2d} \big(-1/n_{\mathfrak{g}}\tau, \mathcal{M}^{^LG}_{\text{Higgs}}(\Sigma_{1})\big)}\\
  %%%%%%%%%%%%%%%%%%%%%%%%%%%%%%%%%%%%%%%%%%%%%%%%%%%%%%%%%%%%%%%%%%%%%%%%%%%%%%%%%%%%%
  \boxed{\substack{\textbf{Topological-Holomorphic}\\\textbf{twist of $\mathcal{N}$=4 SYM} }}
  \arrow[uu, "\mathcal{Q}=\bar{\mathcal{Q}}_1+\bar{\mathcal{Q}}_3" ]
  \arrow[dd, "\mathcal{Q}'=\bar{\mathcal{Q}}_3+t\mathcal{Q}_4"']\\
  %%%%%%%%%%%%%%%%%%%%%%%%%%%%%%%%%%%%%%%%%%%%%%%%%%%%%%%%%%%%%%%%%%%%%%%%%%%%%%%%%%%%%%%
   &\boxed{ \text{CDO}\big(w, \mathcal{M}^{G_{\mathbb{C}}}_{\text{flat}}(\Sigma_{1})\big)}
  \arrow[rrr, leftrightarrow, dashed, "\text{Langlands dual}"]
  &&&\boxed{\text{CDO}\big(w, \mathcal{M}^{{^LG}_{\mathbb{C}}}_{\text{flat}}(\Sigma_{1})\big)}\\ 
  \boxed{\langle \Pi_i\mathcal{O}'_i \rangle_{4d}\big(w, G\big)}
  \arrow[rr, leftrightarrow, dashed, "\text{Langlands dual}"]
  \arrow[ur, leftrightarrow, "\substack{\text{$M_4=\Sigma_{1} \times \Sigma_2$,}\\\text{$\Sigma_1\to 0$}}"']
  \arrow[d, "t=0"']
  &&\boxed{\langle \Pi_i\mathcal{O}'_i \rangle_{4d}\big(w, {^LG}\big)}
  \arrow[urr, leftrightarrow, "\substack{\text{$M_4=\Sigma_{1} \times \Sigma_2$,}\\\text{$\Sigma_1\to 0$}}"']
  \arrow[d,   "t=0"]\\
  \boxed{\langle \Pi_i\mathcal{O}'_i \rangle_{4d,0}\big(w,z,\tau, G\big)}
  \arrow[rr, leftrightarrow, dashed, "\text{Langlands dual}"]
  &&\boxed{\langle \Pi_i\mathcal{O}'_i \rangle_{4d,0}\big(w,z,-1/n_{\mathfrak{g}}\tau, {^LG}\big)}\\
  %\boxed{ \text{CDO}\big(w, \tau, \mathcal{M}^G_{\text{Higgs}}(\Sigma_{1})\big)}
  %\arrow[u, leftrightarrow , "\substack{\text{$M_4=\Sigma_{1} \times \Sigma_2$,}\\\text{$\Sigma_1\to 0$}}"]
  %\arrow[rr, leftrightarrow, dashed, "\substack{\text{Mirror symmetry}\\\text{of Higgs}\\\text{bundles}}"]
  %&&\boxed{\text{CDO} \big(w, -1/n_{\mathfrak{g}}\tau, \mathcal{M}^{^LG}_{\text{Higgs}}(\Sigma_{1})\big)}
  %\arrow[u, leftrightarrow , "\substack{\text{$M_4=\Sigma_{1} \times \Sigma_2$,}\\\text{$\Sigma_1\to 0$}}"']\\
    \end{tikzcd}
\caption{A novel web of mathematical relations stemming from a topological-holomorphic twist of an $\mathcal{N}=4$ supersymmetric gauge theory on $M_4$} \label{fig: web}
\end{center}
\end{figure}
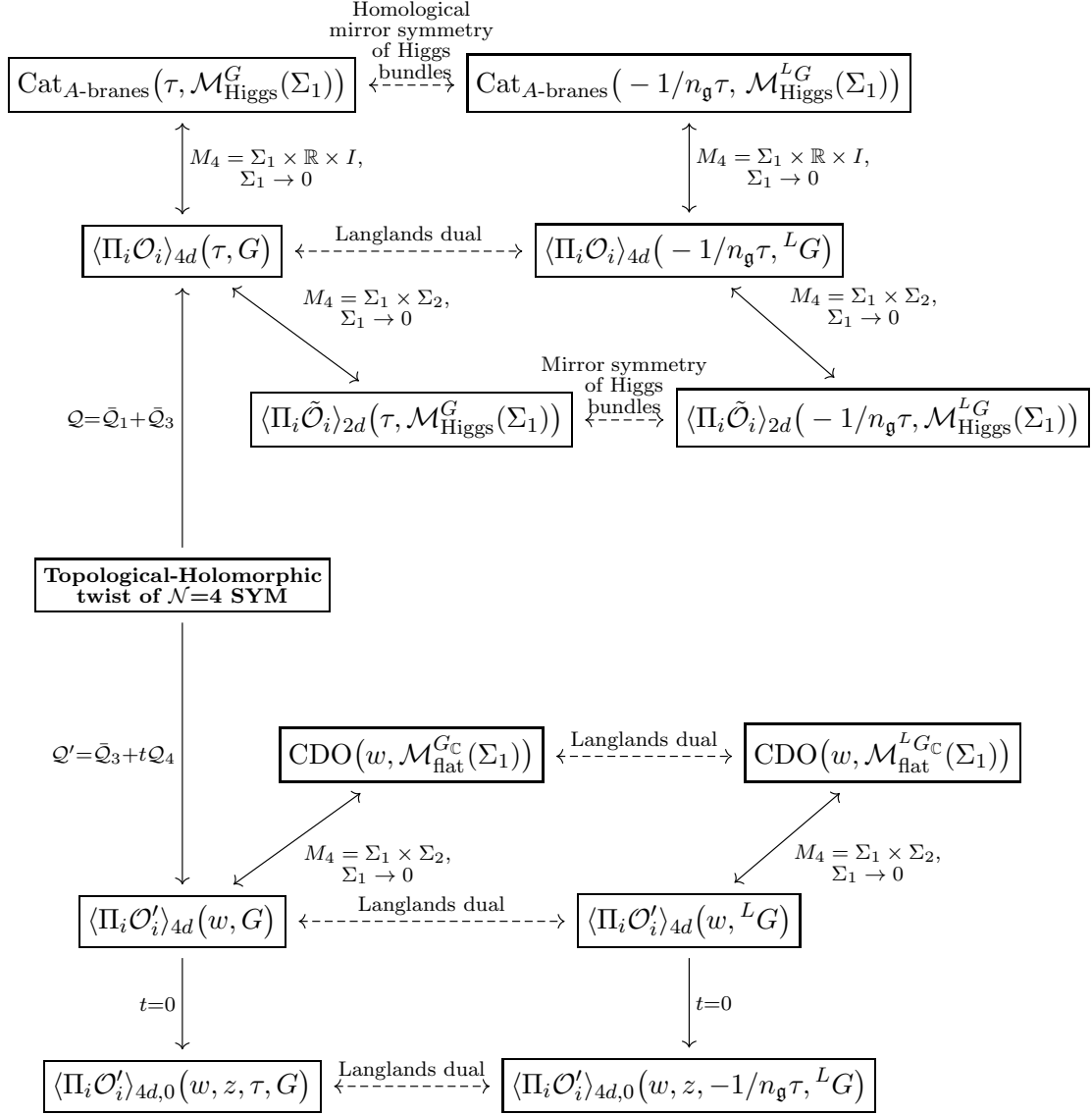

Starting from the topological-holomorphic twist of ${\cal N}=4$ SYM on $M_4 = \Sigma_1 \times \Sigma_2$, via the cohomology of different linear combinations of supercharges and $S$-duality, we will have the dualities \eqref{langlands dual 4d top corr}, \eqref{langlands dual 2d higgs}, \eqref{dual Cat A brane}, \eqref{langlands dual 4d top-hol corr}, \eqref{langlands dual 2d top-hol }, \eqref{langlands 4d hol single}, which
%\eqref{langlands 2d hol single} 
 relate the various mathematical objects that were derived and discussed in this paper. This is depicted in a novel web of mathematical relations in Fig.~\ref{fig: web}.
 %We have also taken the liberty to include the case where we can shrink $\Sigma_1$ when $t=0$ in $\mathcal{Q}'$-cohomology.

\printbibliography
\end{document}